\pdfoutput=1 
\documentclass[preprint,12pt]{elsarticle}

\usepackage{latexsym}
\usepackage{amsmath}
\usepackage{amsfonts}
\usepackage{amssymb}
\usepackage{psfrag}
\usepackage{epsfig}
\usepackage[english]{babel}
\usepackage{color}
\usepackage{soul}
\usepackage{url}

\usepackage[english,ruled,vlined]{algorithm2e}

\usepackage{pgfplots}
\pgfplotsset{compat=newest}
\usetikzlibrary{shapes}
\usetikzlibrary{plotmarks}
\usepackage{float}

\newsavebox{\fmbox}

\newtheorem{theorem}{Theorem}[section]
\newtheorem{remark}[theorem]{Remark}

\newcommand{\ds}{\displaystyle}

\def\Vo{\vbox{\offinterlineskip\hbox{\kern 3pt$\scriptstyle\circ$}
\kern 1pt\hbox{$V$}}}
\def\Ho{\vbox{\offinterlineskip\hbox{\kern 3pt$\scriptstyle\circ$}
\kern 1pt\hbox{$H$}}}
\def\Wo{\vbox{\offinterlineskip\hbox{\kern 3pt$\scriptstyle\circ$}
\kern 1pt\hbox{$W$}}}

\newcommand{\jumpEK}[1]  {[[#1]]_{E,K} }            

\let\vv =\v

\newcommand{\be}{\begin{equation}}
\newcommand{\ee}{\end{equation}}
\newcommand{\beq}{\begin{eqnarray}}
\newcommand{\eeq}{\end{eqnarray}}
\newcommand{\beqs}{\begin{eqnarray*}}
\newcommand{\eeqs}{\end{eqnarray*}}
\newcommand{\bt}{\begin{theorem}}
\newcommand{\et}{\end{theorem}}
\newcommand{\bex}{\begin{example}}
\newcommand{\eex}{\end{example}}
\newcommand{\br}{\begin{remark}}
\newcommand{\er}{\end{remark}}
\newcommand{\bc}{\begin{corollary}}
\newcommand{\ec}{\end{corollary}}
\newcommand{\bl}{\begin{lemma}}
\newcommand{\el}{\end{lemma}}
\newcommand{\bp}{\begin{proposition}}
\newcommand{\ep}{\end{proposition}}
\newcommand{\bd}{\begin{definition}}
\newcommand{\ed}{\end{definition}}
\newcommand{\bas}{\begin{assumption}}
\newcommand{\eas}{\end{assumption}}

\newcommand{\R}{\mathbb{R}}

\newcommand{\PP}{\mathbb{P}}

\newcommand{\caE}{{\cal E}}

\newcommand{\caK}{{\cal K}}

\def\eb{{\bf e}}
\def\u{{\bf u}}

\def\n{{\bf n}}

\def\f{{\bf f}}

\def\vv{{\bf v}}
\def\w{{\bf w}}

\def\z{{\bf z}}

\def\x{{\bf x}}
\def\y{{\bf y}}
\def\V{{\bf V}}

\def\K{{\bf K}}

\def\F{{\bf F}}
\def\0{{\bf 0}}

\newcommand{\nEK}{\n_{E,K}}

\newcommand{\bnabla}{\mbox{\boldmath{$\nabla$}}}

\newcommand{\bs}{\mbox{\boldmath{$\sigma$}}}
\newcommand{\e}{\mbox{\boldmath{$\varepsilon$}}}

\newcommand{\wzh}{\widehat{\mathbf{z}}_{h}}

\begin{document}

\begin{frontmatter}



\title{Quantifying discretization errors for soft-tissue simulation in computer assisted surgery: a preliminary study}


\author[IMM]{Michel Duprez}
\ead{mduprez@math.cnrs.fr}
\author[UL]{St\'ephane Pierre Alain Bordas}
\ead{stephane.bordas@northwestern.edu}
\author[TXS]{Marek Bucki}
\ead{marek.bucki@texisense.com}
\author[LMB]{Huu Phuoc Bui}
\ead{huu-phuoc.bui@math.cnrs.fr}
\author[LMB]{Franz Chouly\corref{cor1}}
\ead{franz.chouly@univ-fcomte.fr}
\author[I3M]{Vanessa Lleras}
\ead{vanessa.lleras@umontpellier.fr}
\author[UTFSM]{Claudio Lobos}
\ead{clobos@inf.utfsm.cl}
\author[LMB]{Alexei Lozinski}
\ead{alexei.lozinski@univ-fcomte.fr}
\author[LBM]{Pierre-Yves Rohan}
\ead{pierre-yves.rohan@ensam.eu}
\author[UL]{Satyendra Tomar}
\ead{tomar.sk@iitkalumni.org}

\cortext[cor1]{Corresponding author}

\address[IMM]{
Institut de Mah\'ematiques de Marseille,
Universit\'e d'Aix-Marseille,
39, rue F. Joliot Curie,
13453 Marseille Cedex 13,
France.
}
\address[UL]{
Facult\'e des Sciences, de la Technologie et de la Communication,
Department of Computational Engineering Sciences,
Universit\'e du Luxembourg,
Maison du Nombre,
6, Avenue de la Fonte
L-4364 Esch-sur-Alzette, Luxembourg.
}
\address[TXS]{
TexiSense,
2, Avenue des Puits,
71300 Montceau-les-Mines, France.
}
\address[LMB]{
Laboratoire de Math\'ematiques de Besan\c{c}on, UMR CNRS 6623,
Universit\'e Bourgogne Franche-Comt\'e,
16, route de Gray,
25030 Besan\c{c}on Cedex,
France.
}
\address[I3M]{
Institut de Math\'ematiques et de Mod\'elisation de Montpellier,
Universit\'e Montpellier,
Case courrier 051,
Place Eug\`ene Bataillon,
34095 Montpellier Cedex, France.
}
\address[UTFSM]{
Departamento de Inform\'atica,
Universidad T\'ecnica Federico Santa Mar\'ia,
Av. Vicu\~{n}a Mackenna 3939,
8940897, San Joaqu\'in, Santiago,
Chile.
}
\address[LBM]{
LBM/Institut de Biom\'ecanique Humaine Georges Charpak, {Arts et Metiers ParisTech}, 151 Boulevard de l'H\^opital, 75013 Paris, France.
}

\begin{abstract}
Errors in biomechanics simulations arise from modeling and discretization. 
Modeling errors are due to the choice of the mathematical model whilst discretization errors measure the impact of the choice of 
the numerical method on the accuracy of the approximated solution to this specific mathematical model.
A major source of discretization errors is mesh generation from medical images, that remains one of the major bottlenecks in the development of reliable, accurate, automatic and efficient personalized, clinically-relevant Finite Element (FE) models in biomechanics. 
The impact of mesh quality and density on the accuracy of the FE solution can be quantified with \emph{a posteriori} error estimates.
Yet, to our knowledge, the relevance of such error estimates 
 for practical biomechanics problems has seldom been addressed, see \cite{bui2017real}.
In this contribution, we propose 
an implementation of some \emph{a posteriori} error estimates 
to quantify the \emph{discretization errors} and {to optimize the  mesh}. 
More precisely, {we focus on error estimation for a user-defined quantity of interest with the Dual Weighted Residual (DWR) technique.}
We test its applicability and relevance in two situations, corresponding to computations for a tongue and an artery, using a simplified setting, \emph{\emph{i.e.}}, plane linearized elasticity with contractility of the soft-tissue modeled as a pre-stress. Our results demonstrate the feasibility of such methodology to estimate the actual solution errors and to reduce them economically through mesh refinement.
\end{abstract}


\end{frontmatter}


\section{Introduction}
\label{sec:intro}

Patient-specific {f}inite {e}lement models of soft tissue and organs have received an increasing amount of interest in the last decades. 
{Such finite element models are widely employed to investigate both, the underlying mechanisms that drive normal physiology of biological soft tissues \cite{avril_anisotropic_2010, franquet_identification_2013, gras_hyper-elastic_2012}, and the mechanical factors that contribute to the onset and development of diseases such as pressure ulcers \cite{loerakker2011effects, shilo_identification_2012}, atherosclerosis or aneurysms \cite{floch_vulnerable_2009, badel_finite_2013, romo_vitro_2014}, or multilevel lumbar degenerative disc diseases \cite{park_effects_2013, schmidt_effect_2012}, to name a few. Finite element models are also valuable tools that contribute to the development of medical devices{, see \emph{\emph{e.g.}}} vascular stent-grafts \cite{perrin_patient-specific_2015}, artificial facet systems for spinal fusion \cite{goel_anatomic_2007} or knee braces \cite{pierrat_evaluation_2014}. They have the potential to improve prevention strategies \cite{luboz_personalized_2017, trabelsi2015patient}, surgical planning \cite{buchaillard_simulations_2007} and pedagogical simulators for medical training \cite{bro-nielsen_real-time_1996, cotin_real-time_1999,buttin_biomechanical_2013,courtecuisse2014real} }.

In this context{,} one major issue is meshing, since the reliability of the predicted {mechanical response} arising from computer simulation {heavily relies} on the quality of the underlying finite element mesh: 
if some  elements of the mesh are too distorted or if the mesh is too coarse in some regions,  
the numerical solution may deteriorate significantly  {\cite{gratsch_posteriori_2005}}.

Whilst generating meshes of complex shapes with tetrahedral elements is generally possible thanks to advanced meshing algorithms, e.g. \cite{geuzaine2009gmsh,remacle2017two}, low-order Lagrange tetrahedral elements are unsuitable for most biomechanical problems due to volumetric locking. To circumvent such issues, strain smoothing approaches were developed \cite{ong2015stability,lee2017strain}, which have the drawback of leading to larger system matrix bandwidth, but the advantage of being easily parallelized on graphical processing units thanks to nodal integration.
Hexahedral low-order Lagrange elements alleviate locking issues, but research on the automatic generation of hexahedral meshes is still ongoing \cite{bommes2013quad,pellerin2017combining,ray2017hexahedral}, spurred by a recrudescent surge in research on polyhedral mesh generation \cite{polymesh_hu2016centroidal,polymesh_mouly2017minimizing,polymesh_budninskiy2016optimal,polymesh_chen2017interface,polymesh_walton2017advances} and approximations such as the virtual finite element method \cite{poly_beirao2016virtual}, hybrid high-order (HHO) methods \cite{dipietro2015}, polyhedral finite elements \cite{poly_talebi2016stress,poly_kraus2017finite,poly_natarajan2017new,poly_francis2017linear}, smoothed finite elements \cite{poly_sellam2017smoothed}, scaled finite elements \cite{poly_talebi2016stress,poly_natarajan2017scaled}, for various applications, including large strain hyper-elasticity \cite{poly_rajagopal2017hyperelastic,abbashal} and optimization \cite{poly_chau2017polytree,poly_chau2017polytree,poly_nguyen2017new}.

The patient-specific mesh has to be built from segmented medical images (CT, MRI, ultra-sound){,} and {has} to conform to anatomical details with potentially complex topologies and geometries {\cite{telfer_simplified_2016,bijar2016atlas,al-dirini_development_2016}}, which led to the design of algorithms that aim to optimize the quality of the generated mesh by reducing the distortion of the elements {\cite{zhang_patient-specific_2007,laville_parametric_2009, shang_hexahedral_2017}}. These algorithms may also have to satisfy a number of additional constraints such as minimizing human intervention (automation), preservation of certain important anatomical details or robustness with respect to data 
\cite{couteau00,alliez05,Ito2009,Shepherd2009,lobos-2010}. 
In general the quality of a given mesh can be assessed through purely geometrical criteria, that allow in some way to quantify the distortion of the geometry of the elements 
and how far they are from their ideal shape \cite{frey-george-2008, bucki-2011, burkhart_finite_2013, george-2016}.

To circumvent or simplify the mesh generation issue, implicit/immersed boundary approaches have been proposed, where the mesh does not conform to the geometry, which is treated by implicit functions such as level sets and enriched finite element methods. This idea was proposed in \cite{impli_moes2003computational} and later generalized in \cite{impli_moumnassi2011finite,moumnassi2014analysis}, \cite{haslinger2009,burman2015}, \cite{impli_ib_liu2016stress,impli_kalameh2016semi} and \cite{impli_rodenas2013creation,impli_tur2015stabilized,impli_fries2017higher}, in 
combination with (goal-oriented) error estimates. Although promising, applications of such approaches to patient-specific geometries remain in their infancy, see a review of related methods in \cite{impli_bordas2010recent}.
Yet another approach consists in directly using the image as a model which could enable simulations without any mesh generation \cite{li2016biomechanical}. In ``image as a model'', the boundary of the domain is smeared, which significantly complicates imposition of boundary conditions, particularly contact.
Finally, meshless methods \cite{nguyen2008meshless} are possible alternatives which may simplify biomechanics simulations by relaxing the constraints posed on mesh quality and simplifying local mesh adaptation. Comparatively to Galerkin meshfree methods \cite{lim2004use,de2005point,lim2007real}, point collocation methods \cite{banihani2009comparison}, also known as generalized finite difference methods, are potentially competitive as they do not require any numerical integration. Quality control of such collocation methods remains an open problem, 
as well as conditioning of the system matrix, which is strongly related to the choice of the stencil \cite{davydov2016optimal}. On the other hand, collocation methods are easily parallelized, for instance on graphical processing units.

Beyond mesh quality, mesh density is another, related, parameter which must be controlled during biomechanics simulations. 
Solutions must be obtained on commodity hardware within clinical time scales: milliseconds (for surgical training);  minutes (for surgical assistance); hours (for surgical planning).
Therefore, and although this would lead to the most accurate solution, it is impractical to use a uniformly fine mesh over the whole domain.
This remark begs the question: ``given a tolerable error level, what is the coarsest possible mesh which will provide the required accuracy.''
This leads to the notion of ``mesh optimality,'' which is achieved for an optimal balance between the accuracy in a given quantity of interest to the user and the associated computational cost.
It is probably intuitively understood that this ``optimality'' criterion, and the resulting optimized mesh both depend on the quantity of interest and that, in general, the optimal mesh will be non-uniform, displaying local refinement around specific regions. 
{
A possible criterion for mesh adaptation can be any {\it a priori} knowledge of the problem or its solution such as geometry,  material properties or boundary layers e.g.  localized loads, contacts, sharp features, material interfaces. Similarly, knowledge of the quantity of interest can help guide local mesh refinement.
{
Nevertheless, such mesh refinement guidelines are generally \emph{ad hoc} and cannot guarantee the resulting mesh will be optimal.


To summarize, the, standard, piecewise linear or quadratic Lagrange-based, finite element method is still the most popular technology to solve biomechanics problems. 
The choice of an optimal mesh, in particular its local refinement level for given problems and quantities of interest remains an open issue. Moreover, without knowing the finite element solution itself, it is practically impossible to quantify the adequacy of a given mesh only from  heuristics or other \emph{ad hoc}  criteria derived from a priori  knowledge of the problem or its exact solution. 

\medskip

{As a result, we aim at addressing the following two questions in this paper:}

\begin{enumerate}

\item For a patient-specific finite element computation, can we provide  some information to the user  about the accuracy of the numerical solution, namely can we compute an approximate  \emph{ discretization error} caused by the choice of the mesh? {{By \emph{discretization error}, we mean the difference between the finite element solution and the exact solution of the same boundary value problem on the same geometry.}}
\item {
Can the numerical solution be used to optimize the mesh in the 
critical regions only, 
to achieve maximum accuracy for a given computational cost, or, conversely, to achieve a given accuracy with a minimum computational cost?}

\end{enumerate}

For the sake of simplicity we do not consider
\begin{enumerate}
\item \emph{modeling errors}, {which arise due to the approximation of the geometry, physical assumptions, and uncertainty on material parameters,}
\item \emph{numerical errors}, {which arise due to linearization, iterative solvers, and machine precision.}
\end{enumerate}

In this paper, we investigate the capability of
\emph{a posteriori} error estimates \cite{ainsworth-oden-2000,verfurth-2013} to provide useful information about the \emph{discretization error}. \emph{A posteriori} error estimates are quantities computed 
from the  numerical solution, that indicate the magnitude of the local error. These estimates are at the core of mesh adaptive techniques \cite{nochetto-2009}. 
Many {\it a posteriori} error 
estimation methods have been 
developed in the numerical analysis community. These methods have different theoretical and practical properties. 
However, despite their great potential, 
error estimates, to the best of our knowledge,  have rarely been considered for  
 patient-specific finite element simulations {in the biomechanical community}. The only reference known to us which addresses discretization error estimation in biomechanics is the very recent paper  \cite{bui2017real} who consider simple but real-time error estimation approaches for needle insertion. 

We limit our study to a simplified setting in order to gain preliminary insights into the behaviour of such \emph{a posteriori} error estimates and to address with the first technical difficulties. We focus on two-dimensional linear elasticity (plane strain) {problems}, with simple boundary conditions (prescribed displacements and tractions), and we assume triangular meshes. This is somehow restrictive in comparison to current practice in soft-tissue simulation. 
Among the existing {\it a posteriori} error estimates{,} we focus on Dual Weighted Residuals (DWR), as presented in, \emph{e.g.},
\cite{becker-rannacher-1996,becker-rannacher-2001} (see {also} \cite{paraschivoiu-peraire-patera-1997,  paraschivoiu-patera-1998, prudhomme-oden-1999, maday-patera-2000,giles-suli-2002,bangerth-rannacher-2003}).
Indeed this method allows to estimate the error for a given quantity of interest. As a matter of fact, for the majority of  applications, controlling the error in the energy norm is not relevant{,} and the error must be controlled for a \emph{specific} quantity of interest to the user ({\emph{e.g.},} stress intensity factors, 
 shear stress or strain intensity at specific locations). 
The DWR method is conveniently implemented in {the} standard finite element library FEniCS \cite{LoggMardal2012} 
and we make use of the implementation described in detail in the paper of Rognes and Logg \cite{rognes-logg-2013}, with minor modifications.

{This paper is organized as follow{s}. In Section \ref{sec:method}, we present the linear elastic problem, the corresponding finite element method, the \textit{a posteriori} error estimates as well as the algorithm for mesh refinement. In Section \ref{sec:results}, we consider two simplified test-cases, inspired by patient-specific biomechanics, where the current methodology is applied. The results are discussed in Section \ref{sec:discussion}.}

\section{Material and methods}\label{sec:method}
\label{sect:mat}

We first present the general problem  consider{ed in this contribution}, that represents a simplified setting for contractile soft-tissue simulation. We then describe in details the computation of the \emph{a posteriori} error estimate: a global estimator {that} provides an estimation of the \emph{discretization error} {and a} local estimator {that drives} the mesh refinement. We end this section with {the description of} a simple algorithm for mesh refinement.

{We first introduce} some useful notations. In what follows, bold letters {such as} $\u,\vv$, indicate vector or tensor valued quantities, while the capital ones (\emph{e.g.}, $\V$, $\K$) represent functional sets involving vector fields. As usual, we denote by $(H^{s}(\cdot))^d$, $s\in \mathbb{R}$, $d=1,2,3$, the Sobolev spaces in one, two or three space dimensions \cite{adams-75}. 
In the sequel the symbol $|\cdot|$  will either denote  the Euclidean norm in $\R^d$, or the 
measure of a domain in $\R^d$.

\subsection{Setting: a ``toy" boundary value problem in linear elasticity}\label{sec:toy problem}

We consider an elastic body whose reference configuration is represented by the domain $\Omega$ in $\R^2$. 
{We consider the plane strain formulation, and allow only small deformations}. 
We suppose that $\partial\Omega$
consists {of} two {disjoint} 
parts $\Gamma_D$ and $\Gamma_N$, with meas$(\Gamma_D) > 0$.
The unit outward normal vector on 
$\partial\Omega$ is denoted {by} $\n$.
A displacement $\u_D=\0$ is applied on $\Gamma_D${, and the} body is subjected {to volume forces $\f \in
(L^2(\Omega))^2$ and} surface loads $\F \in (L^2(\Gamma_N))^2$.
%
%
%
%
We introduce the bilinear form 
\begin{equation*} 
a(\vv,\w) := \int_{\Omega} \bs (\vv) :\e (\w) ~d\x, 
\end{equation*}
which represents the (internal) virtual work associated to passive elastic properties. 
{The notation $\e(\vv) = \frac12 (\bnabla \vv + \bnabla \vv^{^T})/2$ represents the linearized strain tensor field, and
$\bs = (\sigma_{ij})$, $1 \le i,j \le 2$, stands for the stress tensor field, {assumed to be given by Hooke's law}.} 
The linear form 
\begin{equation*} 
l_E(\w) := \int_{\Omega} \f \cdot \w ~d\x + \int_{\Gamma_{N}} \F \cdot \w ~ds
\end{equation*}
stands for the virtual work of external loads in the body and on its surface. 
Finally we represent in a very simplified manner the active properties of soft-tissue as a linear anisotropic pre-stress 
\begin{equation*}\label{eq:defla}
l_A(\w) := - \beta T \int_{\omega_A} \left ( \e (\w) \eb_A \right ) \cdot \eb_A  ~d\x,
\end{equation*}
where $\omega_A$ is the part of the body where muscle fibers are supposed to act, $T \geq 0$ is a scalar which stands for the tension of the fibers, $\eb_A$ is a field of unitary vectors that stands for muscle fibers orientation,
{and} {$\beta \in [0,1]$} is the activation parameter. When $\beta=0$ there is no activation of the muscle fibers, and the value $\beta=1$ corresponds to the maximum activation. This modeling can be viewed as a linearization of some more sophisticated active stress models of contractile tissues (see, \emph{e.g.}, {\cite{cowin2001cardiovascular,payan-ohayon-2017}}).
%
%
%
%
%
%

We want to solve the following weak problem
\begin{equation}\left\{\begin{array}{l}
\textrm{Find a displacement }\u \in \V\textrm{ such that}\\\noalign{\smallskip}
a(\u, \vv) = l(\vv), \quad \forall \, \vv \in \V,
\end{array}\right.\label{eq:primal_weak}
\end{equation}
where $l(\cdot) = l_E(\cdot) + l_A(\cdot)$, and 
where 
$\u$ and 
$\vv$ {lie in} the space of admissible displacements
\begin{align*}
\V := \big \{ \vv\in H^{1}(\Omega)^2 \,|\, ~\vv = \0 \text{~on~} \Gamma_{D} \big \}.
%
\end{align*}


From the displacement field, we are interested in computing a linear  quantity
\begin{equation}\label{def:J}
J: \V \ni \u \mapsto J(\u) \in \R, 
\end{equation}
which can be defined according to a specific application and  the interest of each practitioner. {Thereby, the quantity $J$ will be aptly called {\emph{quantity of interest (QoI)}}.} We will provide its expression(s) for each test case.

\subsection{Finite element method}\label{sec:finite element method}

Consider a family of meshes $(\caK_h)_{h>0}$ 
constituted of triangles and assumed to be subordinated to the decomposition of the boundary 
$\partial\Omega$ into $\Gamma_D$ {and} $\Gamma_N$. 
%
For a mesh $\caK_h$, {we} denote by $\caE_h$ the set of edges, 
{by} $\caE^{int}_h:=\{E\in \caE_h:E\subset \Omega\}$ the set of interior edges,
and {by} $\caE^N_h:=\{E\in \caE_h:E\subset \Gamma_N\}$ the set of boundary edges that correspond to Neumann conditions {(we assume that any boundary edge is either inside $\Gamma_N$ or inside $\Gamma_D$)}.
For an element $K$ of $\mathcal{K}_h$, we set $\caE_K$ the set of edges of $K$, 
$\caE_K^{int}:=\caE_K\cap \caE_h^{int}$ and  $\caE_K^{N}:=\caE_K\cap \caE_h^{N}$. {We also assume that each element $K$ is either completely inside 
$\omega_A$ or completely outside it.}
Let $\bs$ be a second-order tensorial field in $\Omega$, {which is assumed to be} piecewise continuous.
We define the jump of $\bs$ across an interior edge $E$ 
of an element $K$, 
at a point $\y\in E$, as follow{s}
\begin{equation*}
\jumpEK{\sigma} (\y):=\lim\limits_{\alpha\rightarrow 0^+}
\left (\bs(\y+\alpha\nEK) - \bs(\y-\alpha\nEK) \right)  \nEK,
\end{equation*}
where $\nEK$ is the unit normal vector to $E$, pointing out of $K$.

The finite element space $\V_{h}\subset\V$
is built upon
continuous Lagrange finite elements of degree $k=1,2$ 
(see, \emph{e.g.}, \cite{ern_guermond2004}), 
{\it i.e.}
\begin{equation*}
\V_{h}:=\left\{\vv_{h}\in(\mathcal{C}^0(\overline{\Omega}))^d:\vv_{h|K}\in(\PP_k(K))^d,
\forall K\in\mathcal{K}_h,\vv_{h}=\0\mbox{ on }\Gamma_D\right\}.
\end{equation*}
Problem \eqref{eq:primal_weak} is approximated by
\begin{equation}
\left\{\begin{array}{l}
\textrm{Find }\u_{h} \in \V_{h}\mbox{ such that }\\\noalign{\smallskip}
a(\u_{h}, \vv_{h}) = l(\vv_{h}), \qquad \forall \vv_{h} \in \V_{h}.
\end{array}\right.
\label{eq:primal_weak_disc}
\end{equation}



\subsection{Goal-oriented error estimates}
  
We compute goal-oriented error estimates using the  
Dual Weighted Residual (DWR) technique \cite{becker-rannacher-1996,becker-rannacher-2001} (see {also} \cite{bangerth-rannacher-2003, giles-suli-2002, maday-patera-2000, paraschivoiu-patera-1998, paraschivoiu-peraire-patera-1997, prudhomme-oden-1999}), which is inspired from duality arguments (see, \emph{e.g.}, \cite{eriksson-1995}).
We follow the framework described in 
\cite{rognes-logg-2013}, with some minor changes and adaptations. 
%
%

Let us consider $\u_h$ the solution to Problem \eqref{eq:primal_weak_disc}.
{The {weak residual} 
is defined for all $\vv\in\V$ by}
%
\begin{equation*}\label{eq:def residual}
r(\vv) := l(\vv) - a(\u_h, \vv).
\end{equation*}
%
%

%
%

%
%
Let $\z$ denote the solution to the dual problem:
\begin{equation}
\left\{\begin{array}{l}
\mbox{Find }\z \in \V\mbox{ such that}\\\noalign{\smallskip}
a(\vv, \z) = J(\vv), \qquad \forall\, \vv \in \V.
\end{array}\right.
\label{eq:DefDual}
\end{equation}
%
%

%
%
%
The DWR method, in a linear setting, relies on the fundamental observation that
\begin{align}
J(\u) - J(\u_h) = a(\u, \z) - a(\u_h, \z) = l(\z) - a(\u_h, \z) = r(\z).
\label{eq:J_resid}
\end{align}
From this, we design an error estimator of $J(\u) - J(\u_h)$ 
as an approximation of the residual $r(\z)$. We detail the different steps below.

\subsubsection{Numerical approximation of the dual problem and global estimator}\label{sec:dual}
The exact solution $\z$ to the dual system \eqref{eq:DefDual} is unknown in most of {the} practical situations, and thus needs to be approximated.
Let us consider a finite element space $\widehat{\V}_h \subset \V$. {This space is assumed to be} finer than $\V_h$, for instance, made of continuous piecewise polynomials of order $k+1$. 
{The approximation $\widehat{\z}_{h}$ of the solution to the dual problem $\z$ is obtained by solving the following approximate dual problem}
\begin{equation}
\left\{\begin{array}{l}
\mbox{Find }\widehat{\mathbf{z}}_{h} \in \widehat{\V}_{h}\mbox{ such that}\\\noalign{\smallskip}
a(\widehat{\vv}_{h}, \widehat{\mathbf{z}}_{h}) = J(\widehat{\vv}_{h}), \qquad \forall\, \widehat{\vv}_{h} \in \widehat{\V}_h.
\end{array}\right.
\label{eq:DefDual discr1}
\end{equation}


We define
\begin{equation} \label{eq:globalestimatorh}
\eta_h := |r(\wzh)| 
\end{equation}
as the \emph{global estimator} that approximates the residual $r(\z)$. 

\subsubsection{Derivation of local estimators}\label{sec:local est}

Following \cite{becker-rannacher-2001,rognes-logg-2013}, we provide a \textit{local estimator} of the error $|J(\u)-J(\u_h)|$, that can be written in a general form
\begin{equation}\label{estim j(u)-j(u_h)}
\sum\limits_{K\in\caK_h} \eta_K,
\quad
\eta_K := \left | 
\ds \int_K R_K \cdot (\wzh - i_h \wzh) d\x
\, + \sum_{E \in \caE_K} 
\int_E R_{E,K} \cdot (\wzh^i-i_h\wzh ) ds \, 
\right |, \quad \forall K \in \caK_h,
\end{equation}
where the notation $i_h$ stands for the Lagrange interpolant onto $\V_h$. 
The local element-wise and edge-wise residuals are given explicitly by 
$$R_K:=\f_K +{\bf div}\, \bs_A (\u_h)$$ and 
\begin{equation*} 
R_{E,K} :=\begin{cases}
-\dfrac{1}{2} 
\jumpEK{\bs_A(\u_h)}
& \text{if}~ E \in \caE^{int}_K, \\
\F_E-\bs_A( \u_h ) \nEK  & \text{if}~ E \in \caE^N_K, 
\end{cases}
\end{equation*}
where $$\bs_A := \bs(\u^h) + \beta T ( \eb_A \otimes \eb_A ) {\chi_A}.$$
The notation $\chi_A$ stands for the indicator function of $\omega_A$, \emph{i.e.} $\chi_A=1$ in $\omega_A$ and $\chi_A=0$ elsewhere. The quantity $\bs_A$ represents the sum of passive and active contributions within the stress field.
The quantity $\f_K$ (resp. $\F_E$) is a computable approximation of $\f$ (resp. $\F$). 

{The following bound} always hold{s}
\[
\eta_h \leq \sum\limits_{K\in\caK_h} \eta_K,
\]
since compensation effects (balance between positive and negative local contributions) can occur for $\eta_h$, see, \emph{e.g.}, \cite{nochetto-veeser-verani-2009}. Thus $\eta_h$ is expected to be sharper than $\sum\limits_{K\in\caK_h} \eta_K$. In practice,$\sum\limits_{K\in\caK_h} \eta_K$ aims at quantifying the local errors for mesh refinement.

\begin{remark}
Each local estimator $\eta_K$ is made up of two contributions. On one hand, the residuals $R_K$ and $R_{E,K}$ represent the local error in the natural norm. On the other hand, the contribution $(\wzh^i-i_h\wzh)$ coming from the dual problem can be interpreted as a \emph{weight} (or a sensitivity factor) that measures the local impact on the quantity of interest $J(\cdot)$, see, \emph{e.g.}, \cite[Remark 3.1]{becker-rannacher-2001}.
\label{rmq:weight}\end{remark}


\begin{remark}
In \cite{rognes-logg-2013} the local residuals $R_K$ and $R_{E,K}$ are computed implicitly through local problems, in a generic fashion. No significant difference has been observed numerically between their technique and an explicit computation.
\end{remark}

\begin{remark}
{
We have chosen to {compute $\wzh$
through the approximate dual system computed in 
$\widehat{\V}_h \subset \V$ (\emph{i.e.} the space made of continuous piecewise polynomials of order $k+1$)}.
Other strategies are possible: see, \emph{e.g.}, \cite[Section 5.1]{becker-rannacher-2001} for a discussion. For example, the authors of \cite{rognes-logg-2013} {use}
extrapolation of the approximate dual system computed in $\V_h$. 
{We can also mention} \cite{BET11}, where the weight is estimated 
using a residual \textit{a posteriori} error estimate for the dual system, approximated in $\V_h$. The aforementioned techniques are cheaper since the same space is used for the primal and dual solutions, but they can be less accurate.
}
\end{remark}

\subsection{Algorithm for goal-oriented mesh refinement}\label{sec:algo}

In the last sections, we have described the different steps to construct the global and local error estimators. 
Using the D\"orfler marking strategy \cite{dorfler1996}, we now describe{, in Algorithm \ref{algo 1},} a simple algorithm to refine the mesh by taking into account these quantities. In this algorithm, there are two independent numerical parameters: first a parameter $0 < \alpha \leq 1$ 
that controls the level of refinement in D\"orfler marking, and then a tolerance threshold $\varepsilon > 0$ for the global estimator, that serves as a stopping criterion.

\begin{algorithm}[!h]
\caption{Refinement algorithm}
\RestyleAlgo{tworuled}\vspace*{2.5mm}
\SetKwProg{init}{Initialization}{ :}{}\vspace*{1.5mm}
\init{}{\vspace{3mm}
Select an initial triangulation $\mathcal{K}_h$ of the domain $\Omega$.\\\vspace*{1.5mm}
Build the finite elements spaces $\V_{h}$  and $\widehat{\V}_h$. 
}\vspace*{2.5mm}
\SetKwProg{tant}{While}{ do}{Fin tant que}
\tant{$\eta_h > 
\epsilon$}{
\medskip
\begin{enumerate}
\item Compute $\u_{h}\in \V_{h}$ 
 : $a(\u_{h},\vv_{h})={l}(\vv_h), \quad \forall \vv_{h}\in\V_{h}$.
\item Compute $\widehat{\z}_{h}\in \widehat{\V}_{h}$ : 
$a(\widehat{\vv}_{h},\widehat{\z}_{h})=J(\widehat{\vv_h}), \quad \forall \widehat{\vv}_{h}\in\widehat{\V}_{h}$.
\item Evaluate the global error estimator $\eta_h=|r(\wzh)|$.

\item If $\eta_h \leq \varepsilon$, then stop.

\item Evaluate the local estimators
\begin{equation*}
\eta_K := \left | 
\ds \int_K R_K {\cdot} (\wzh - i_h \wzh) d\x
\, + \sum_{E \in \caE_K} 
\int_E R_{E,K} {\cdot} (\wzh^i-i_h\wzh ) ds \, 
\right |, \quad \forall K \in \caK_h.
\end{equation*}

\item Sort the cells $\{K_1,...,K_N\}$ by decreasing order of $\eta_K$.

\item D\"orfler marking:
mark the first $M^*$
cells for refinement where 
$$M^* := \min \left \{ M \in \mathbb{N} \:\middle |\: \sum\limits_{i=1}^{M}\eta_{K_i}\geq\alpha\sum_{K\in\mathcal{K}_h}\eta_K \right \}.$$ 
\item Refine all cells marked for refinement

(and propagate refinement to avoid hanging nodes).
\item Update correspondingly the finite element spaces $\V_{h}$  and $\widehat{\V}_h$.
\end{enumerate}
} \label{algo 1}
\end{algorithm}


\section{Results}\label{sec:results}

We present numerical results for two different test cases: {the biomechanical response of both a human tongue and an artery 
predicted using finite element analysis,
} inspired from studies \cite{bijar2016atlas} and \cite{floch_vulnerable_2009}, respectively.
We propose to assess the discretization error for the two quantities of interest
\begin{equation}\label{def:quantities}
J_1(\u):=\displaystyle\int_{\omega} ( u_x+u_y ) \, d\x
~~~\mbox{ and }~~~
J_2(\u):= 
\displaystyle\int_{\omega} \mbox{div}\: \u \,\, d\x,
\end{equation}
where $u_x$ and $u_y$ are the two components of $\u$ in a Cartesian basis. 
{The first quantity }{$J_1(\u)$ is physically 
related to the displacement in the region of interest $\omega \subset \Omega$. This corresponds to a quantity that can easily be measured experimentally and that is therefore of practical interest. The second quantity $J_2(\u)$ physically corresponds to the internal strain $I_1 = \mbox{tr} (\e(\u))$. This is also of practical interest because many of the mechanisms driving the onset of pathologies are related to shear strains or principal strains.}
The region of interest $\omega$ will be specified in both situations.
All the simulations of this section are performed with
Lagrange finite elements of degree $k=2$,
and the space $\widehat{\V}_h$ in which $\widehat{\z}_h$ is computed
is built from Lagrange finite elements of degree $k=3$.
In Algorithm \ref{algo 1}, the parameter $\alpha$ for D\"orfler marking is fixed at $0.8$, and the stopping criterion $\varepsilon$ will be specified for each application. In the following, the exact value of $J(\u)$ is unknown but is estimated using computations on a very fine uniform mesh.

\subsection{Human tongue with fiber activation}\label{sec:tongue}

{In the first example, we focus on the case study for the activation of the posterior genio-glossus (GGp), that is a lingual muscle located at the root of the tongue and inserted in the front to
the mandible. The activation of this muscle compresses the tongue in the lower part and generates a forward and upward movement of the tongue body, because of the incompressibility of tongue tissues, for example during the production of the phonemes /i/ or /s/. 
} 
The {2D mesh used in this example has been derived from the generic 3D mesh} 
presented in \cite{bijar2016atlas} where the authors developed a process to generate subject-specific meshes. More precisely an automatic atlas-based method was proposed that generates subject-specific meshes \emph{via} a registration guided by Magnetic Resonance Imaging.
The domain $\Omega$ 
is depicted in Figure \ref{fig:tongue medical img} (left). 
{The width and height of the tongue are respectively equal to $73.8$ mm and $53.7$ mm.
For the passive tissue material properties, we use {the values reported {in} \cite{gerard_non-linear_2005} based on indentation experiments on a cadaver’s tongue. The authors initially proposed an incompressible two parameter Yeoh hyperelastic material model and fitted the material constants to the data. In this work, a linear elastic material model is assumed. According to \cite{tracqui_transmission_2004}, linearisation of the model proposed {in} \cite{gerard_non-linear_2005} yields $E$ $\simeq$ $6c10$ $=0.6$ MPa. {For the sake of simplicity} Poisson ratio is assumed to be $\nu=0.4$.} {No volumic force field is applied: $\f=\0$.} 
The direction of the fibers $\eb_A$ is 
depicted in Figure \ref{fig:tongue medical img} (center) and corresponds approximately to the posterior genioglossus muscle {\cite{bijar2016atlas}.}
{Other parameters for fiber activation have been chosen {as $T$ = 2e-5 {MPa}} and $\beta=1$}. 
The tongue is attached to the hyoid bone and to the mandible, {which} are supposed {to be} fixed. {This} leads to a homogeneous Dirichlet boundary condition such as  depicted in Figure \ref{fig:tongue medical img} (right).
On the remaining part of the boundary a homogeneous Neumann condition ($\F=\0$) is applied.
The orange part depicts the region $\omega_A$ where fibers are supposed to be located.  The green part depicts the region of interest $\omega$ for the computation of $J_1$ and $J_2$. 

\begin{figure}[!h]
\begin{center}
\includegraphics[scale=1.4]{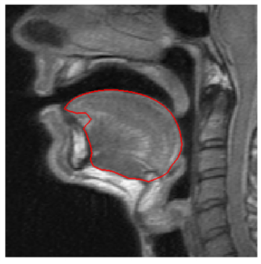} 
\includegraphics[scale=0.11]{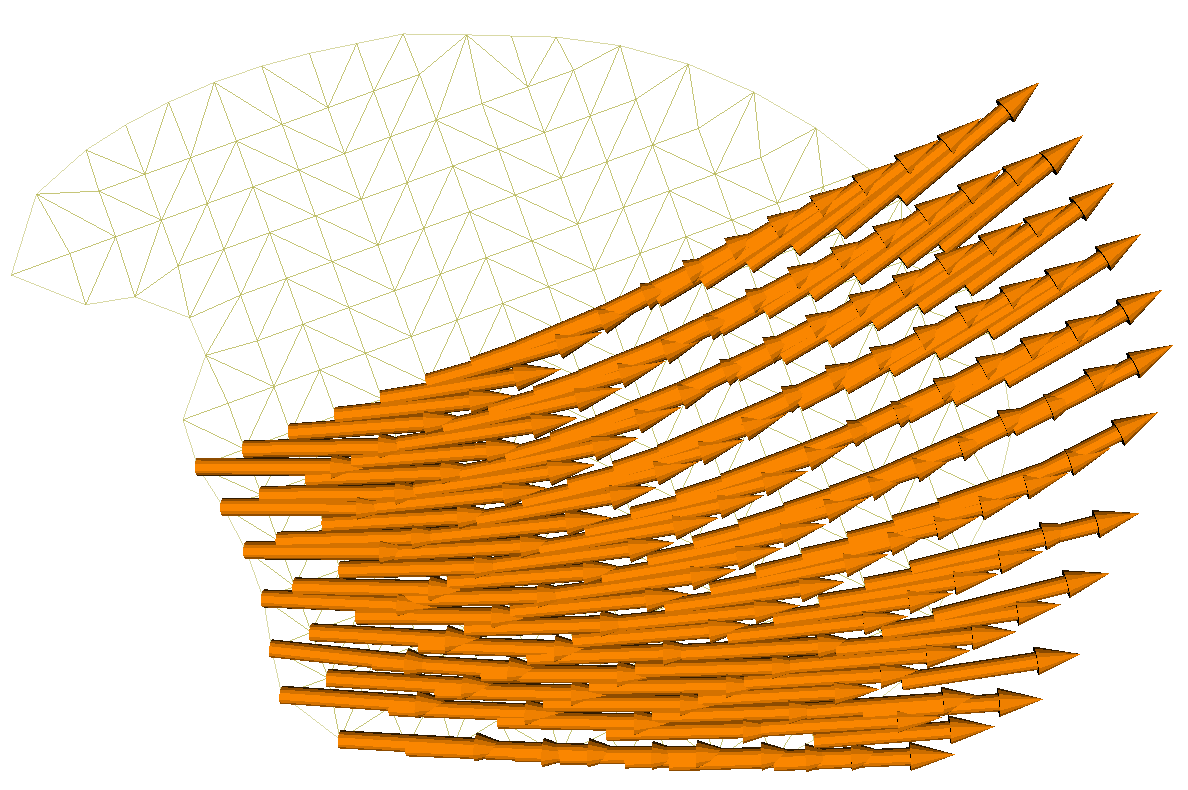} 
\includegraphics[scale=0.35]{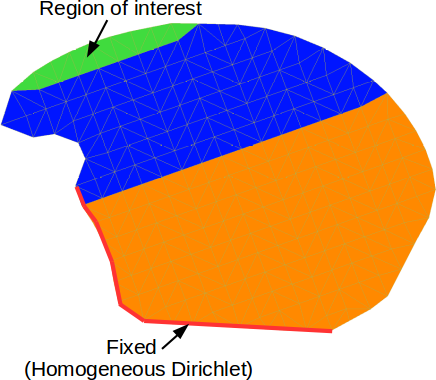} 
\end{center}
\caption{Tongue model: initial geometry (left), fiber orientation (center) and  region of interest (right).}
\label{fig:tongue medical img}\end{figure}

{The resulting displacement is depicted in Figure \ref{fig:tongue dual} (left). 
We computed the relative displacement and the strain intensity, which maximal values are of 5.7 $\%$ and 4.8 $\%$, respectively: thus the small displacement and small strain assumptions are both verified in this case.
The parameter $T$ has been chosen accordingly in order to respect these assumptions.}
In Figure \ref{fig:tongue dual}, the dual solutions  for the quantities of interest $J_1$ (center) and $J_2$ (right) are represented. As mentioned in Remark \ref{rmq:weight}, the dual solution $z$ is used as a weight in the computation of the estimators, and influences the local refinement. 


\begin{figure}[!h]
\hfill\includegraphics[scale=0.08]{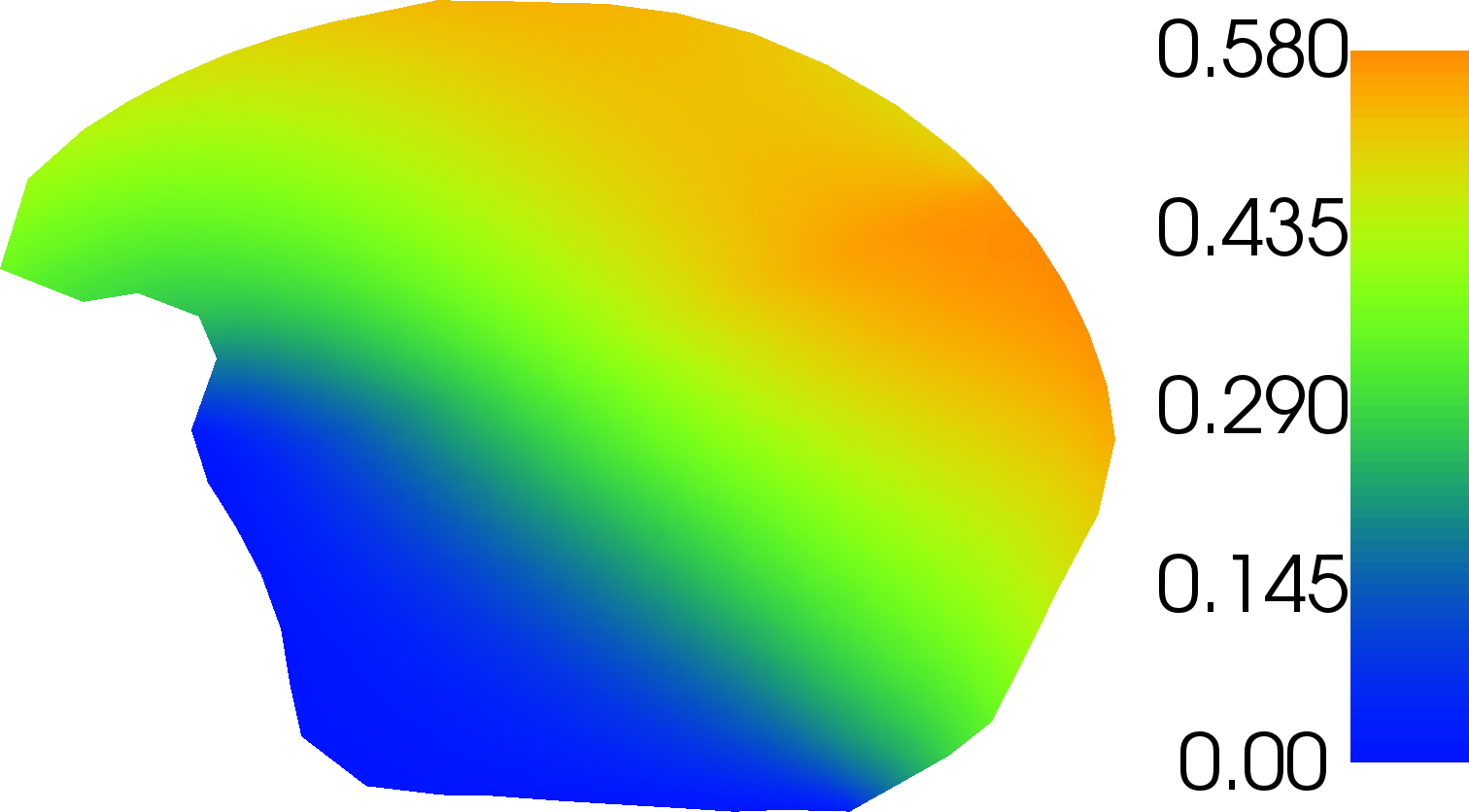}\hfill
\includegraphics[scale=0.08]{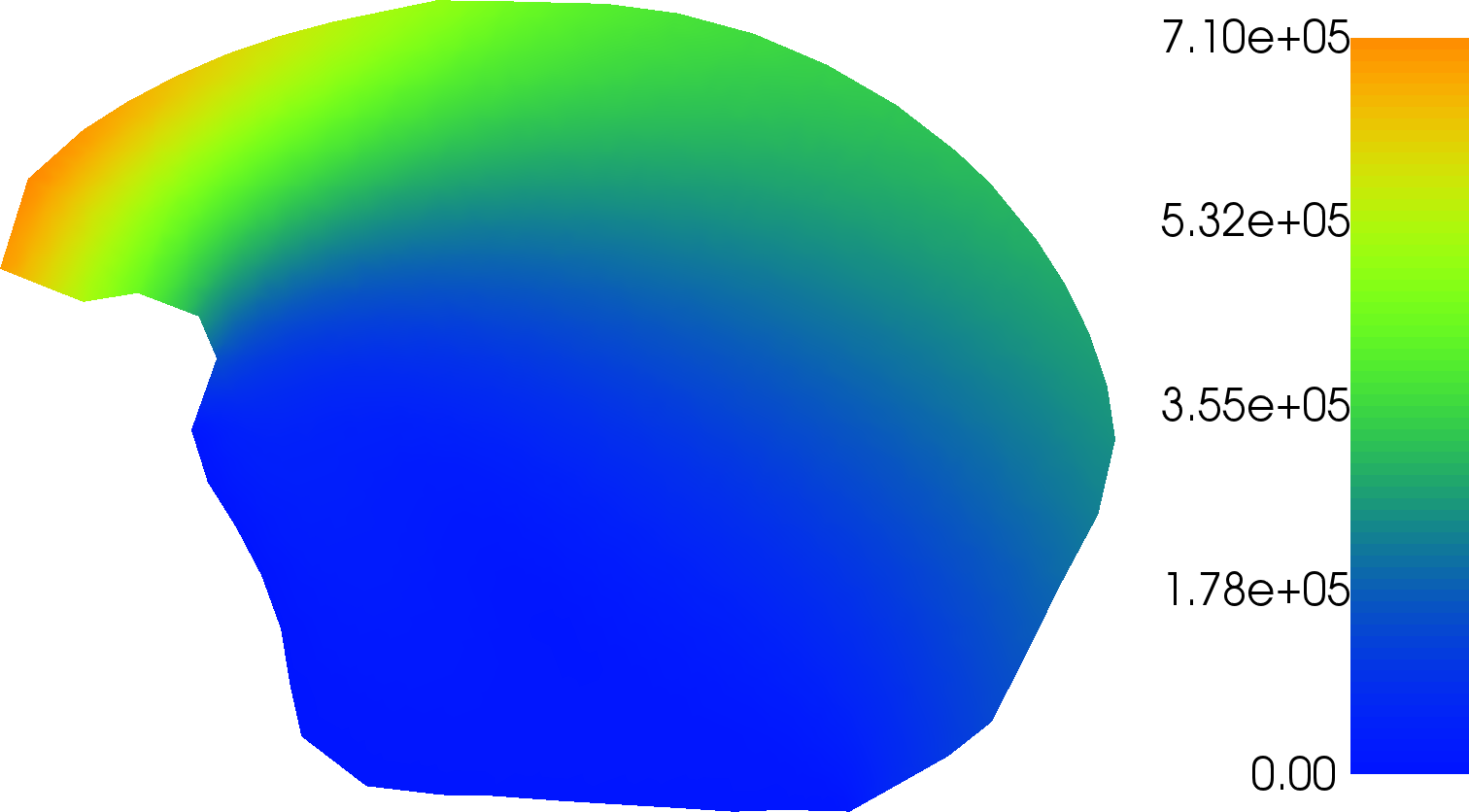}\hfill
\includegraphics[scale=0.08]{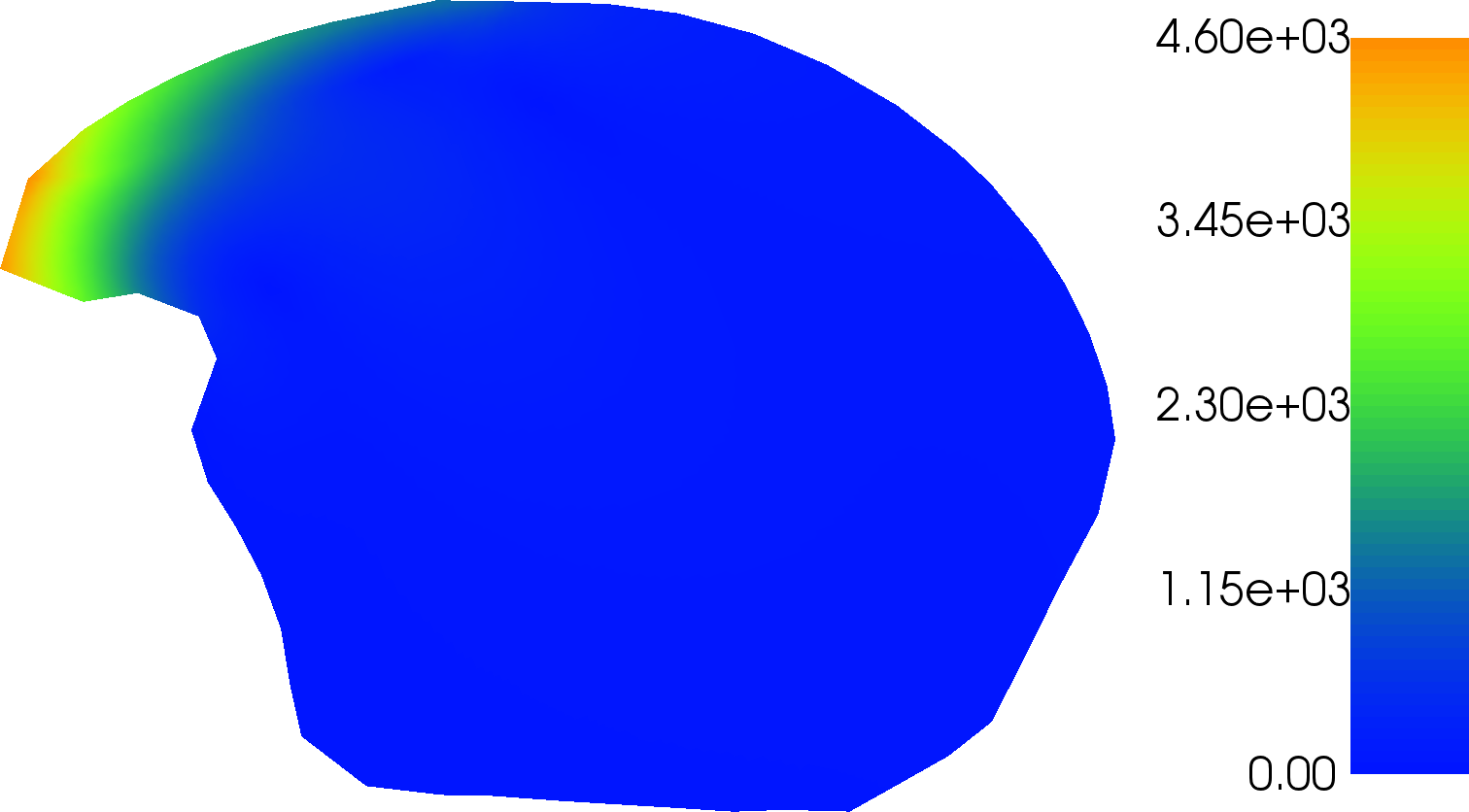}\hfill
\caption{Tongue model: displacement (left), dual solutions  for $J_1$ (center) and for $J_2$ (right).}
\label{fig:tongue dual}\end{figure}


{
{
We present the final mesh after 2 and 8 iterations of Algorithm \ref{algo 1} for both quantities of interest $J_1$ and $J_2$, in Figure \ref{fig:mesh tongue 1} and Figure \ref{fig:mesh tongue 2}, respectively. We first remark that the refinement occurs in some specific regions such as {those} near Dirichlet-Neumann transitions and concavities on the boundary. Note as well that the refinement is stronger for $J_2$ at the boundary of the region of interest $\omega$,.}
}

\begin{figure}[!h]
\begin{center}
\includegraphics[scale=0.119]{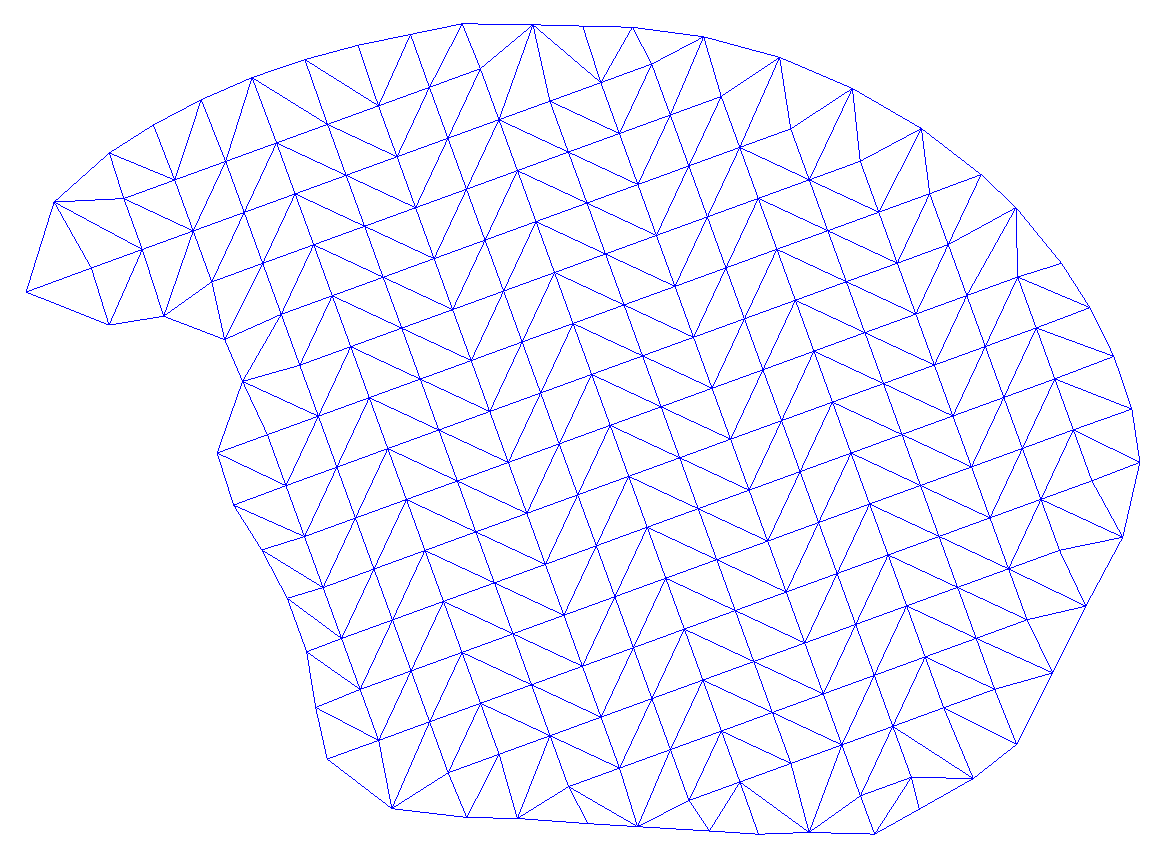} 
\includegraphics[scale=0.119]{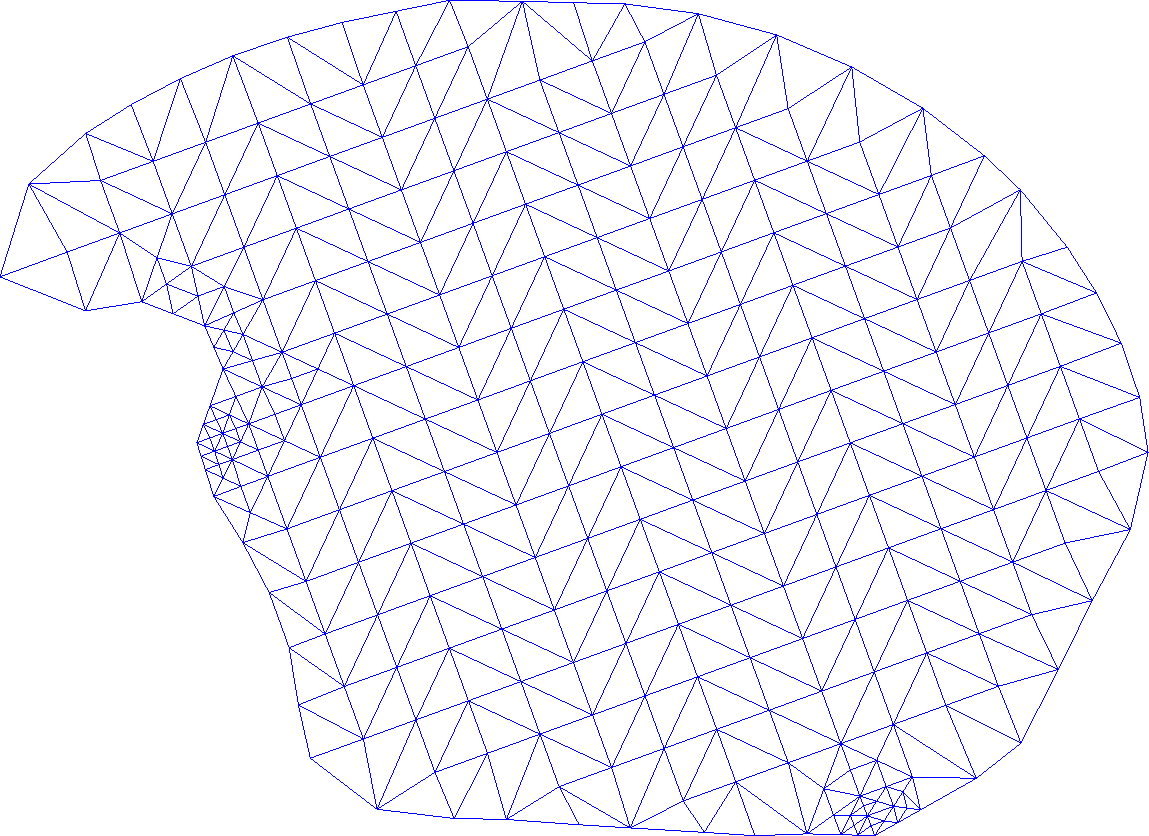}
\includegraphics[scale=0.119]{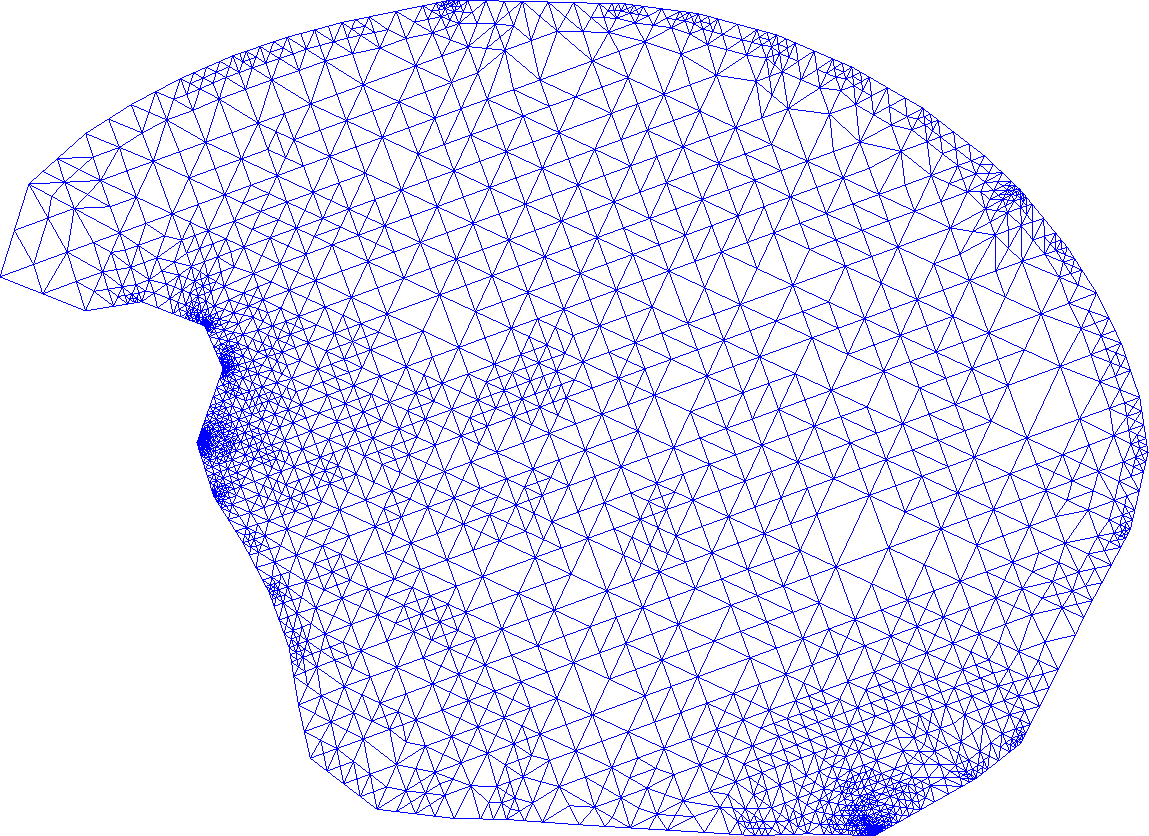}
\end{center}
\caption{ Tongue mesh: refinement driven by the QoI $J_1$.
Initial mesh (left) with 426 cells and a relative error of 1.07e-2, adapted meshes after 2 iterations (center) with 523 cells and a relative error of 2.82e-3 and after 8 iterations (right) with 5143 cells and a relative error of 3.82e-05. 
}
\label{fig:mesh tongue 1}
\end{figure}

\begin{figure}[!h]
\begin{center}
\includegraphics[scale=0.119]{tongue_initial_mesh.png}
\includegraphics[scale=0.119]{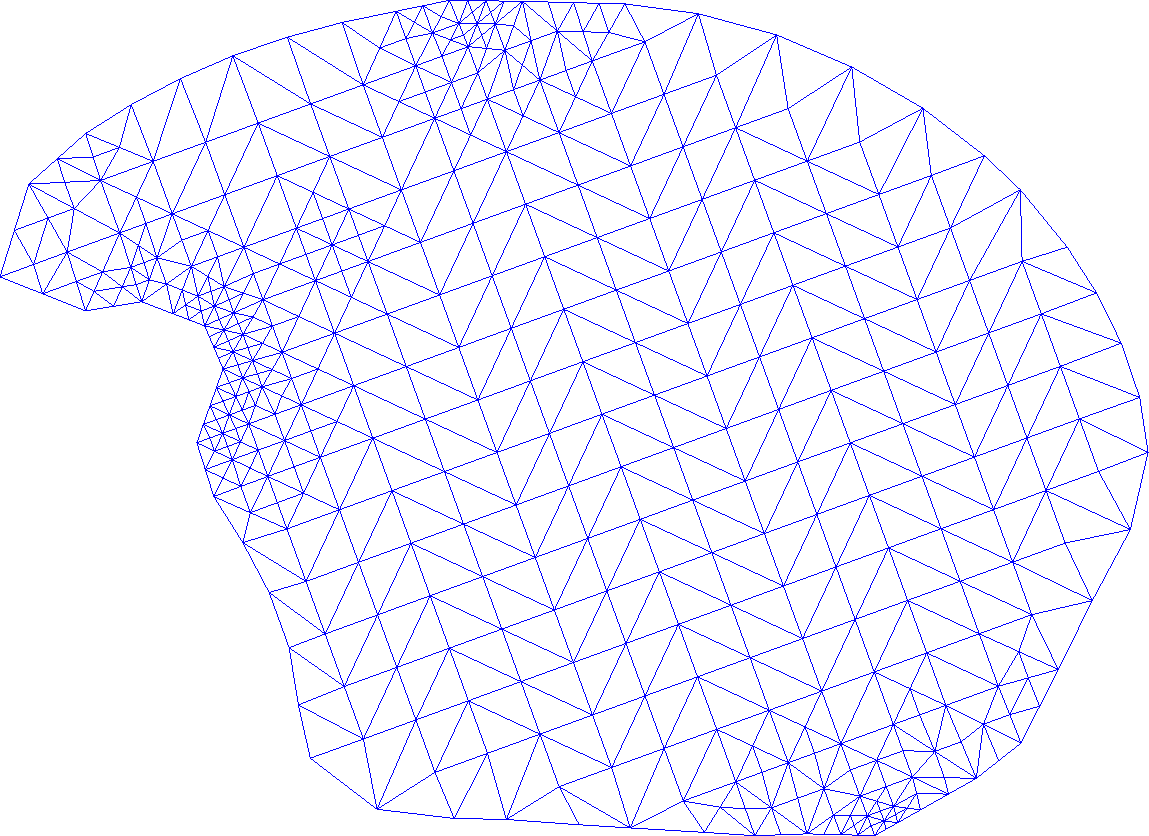}
\includegraphics[scale=0.119]{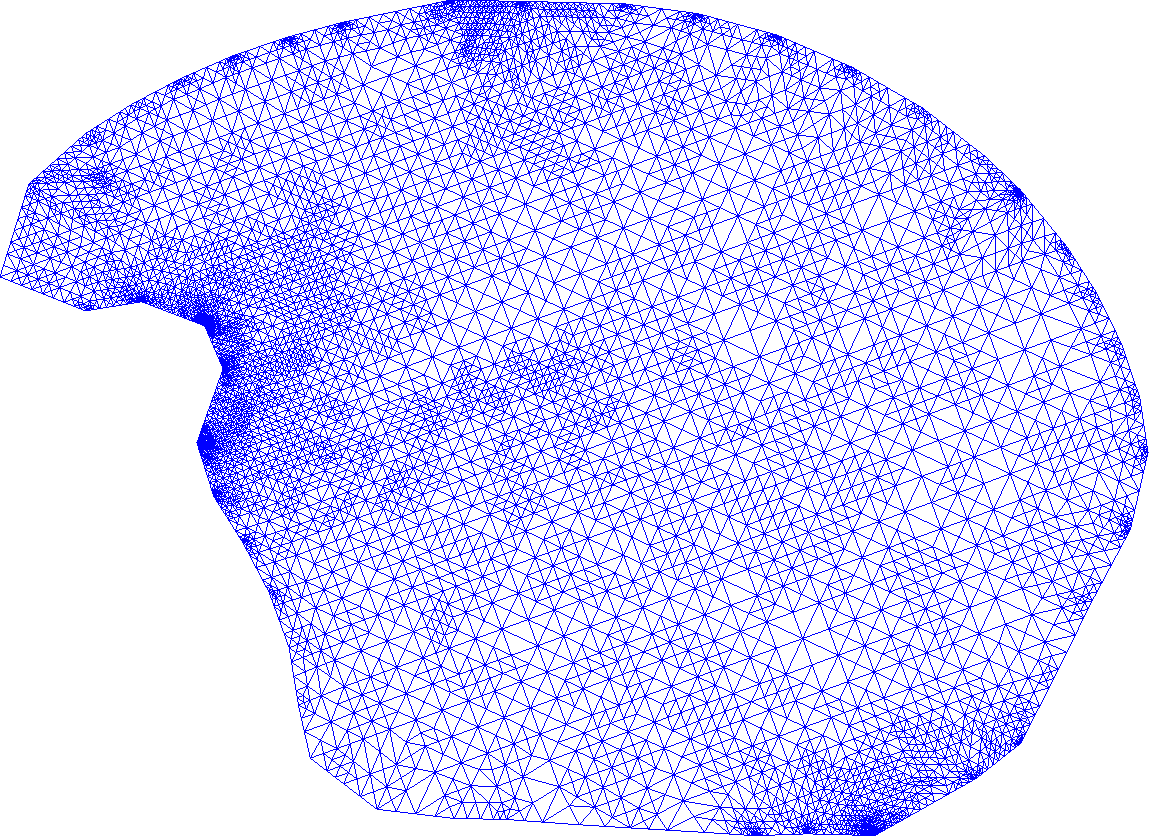}
\end{center}
\caption{ 
Tongue mesh: refinement driven by the QoI $J_2$.
Initial mesh (left) with 426 cells and a relative error of 2.51e-2, adapted meshes after 2 iterations (center) with 766 cells and a relative error of 2.21e-3 and after 8 iterations (right)
 with 13513 cells and a relative error of 2.44e-5.}
\label{fig:mesh tongue 2}
\end{figure}

%
%

{Figure \ref{fig:tongue graph} depicts the relative goal-oriented errors 
$|J_1(\u) - J_1(\u_h)|/|J_1(\u)|$ (left) and $|J_2(\u) - J_2(\u_h)|/|J_2(\u)|$ (right) versus $N$, the number of cells of the mesh, both for uniform refinement (blue) and adaptive refinement (red). The stopping criterion $\varepsilon$ has been fixed { to 2e-4 and 1e-6,} respectively. In each situation, we observe that, as expected, adaptive refinement performs better: not only it leads to a lower error but it also converges much faster when the number of cells $N$ is increased.
}

\begin{figure}[H] 
\begin{center}
\begin{tikzpicture}[thick,scale=0.85, every node/.style={scale=1.0}] \begin{loglogaxis}[xlabel=N,ylabel=$|J_1(u)-J_1(u_h)|/|J_1(u)|$,
xmin=2e2,xmax=3e4
,legend pos=south west, legend columns=1]
 \addplot[color=blue,mark=*] coordinates { 
(426.0,0.0107454970656)
(1704.0,0.00558663947473)
(6816.0,0.00270497706716)
 };
  \addplot[color=red,mark=*]coordinates { 
(426.0,0.0107454970656)
(469.0,0.00568626098031)
(523.0,0.00282549659388)
(619.0,0.00143693184573)
(905.0,0.000705423898089)
(1394.0,0.000343784427098)
(2227.0,0.000159798699974)
(3309.0,8.0952870149e-05)
(5143.0,3.81894124884e-05)
(8106.0,1.81534044685e-05)
(12452.0,8.64993143884e-06)
(18162.0,4.16789243563e-06)
 };
 \legend{ Uniform refinement, Adaptive refinement}
\end{loglogaxis} 
\end{tikzpicture} 
\begin{tikzpicture}[thick,scale=0.85, every node/.style={scale=1.0}] \begin{loglogaxis}[xlabel=N,ylabel=$|J_2(u)-J_2(u_h)|/|J_2(u)|$,
xmin=2e2,xmax=2e4
,legend pos=south west, legend columns=1]
 \addplot[color=blue,mark=*] coordinates { 
(426.0,0.0240559117099)
(1704.0,0.00497735990676)
(6816.0,0.00193601804228)
 };
  \addplot[color=red,mark=*]coordinates { 
(426.0,0.0240559117099)
(509.0,0.00554448877477)
(766.0,0.00221056123228)
(1274.0,0.00109093152754)
(2054.0,0.000488616640837)
(3436.0,0.000243170269827)
(5484.0,0.000108724666567)
(8674.0,5.04921146455e-05)
(13513.0,2.4360089441e-05)
 };
 \legend{ Uniform refinement, Adaptive refinement}
\end{loglogaxis} 
\end{tikzpicture} 
\end{center}
\caption{Tongue model: relative error for the QoI $J_1$ (left) and $J_2$ (right) \emph{vs.} the number $N$ of cells in the case of uniform (blue) and adaptive (red) refinement.}\label{fig:tongue graph}
\end{figure}
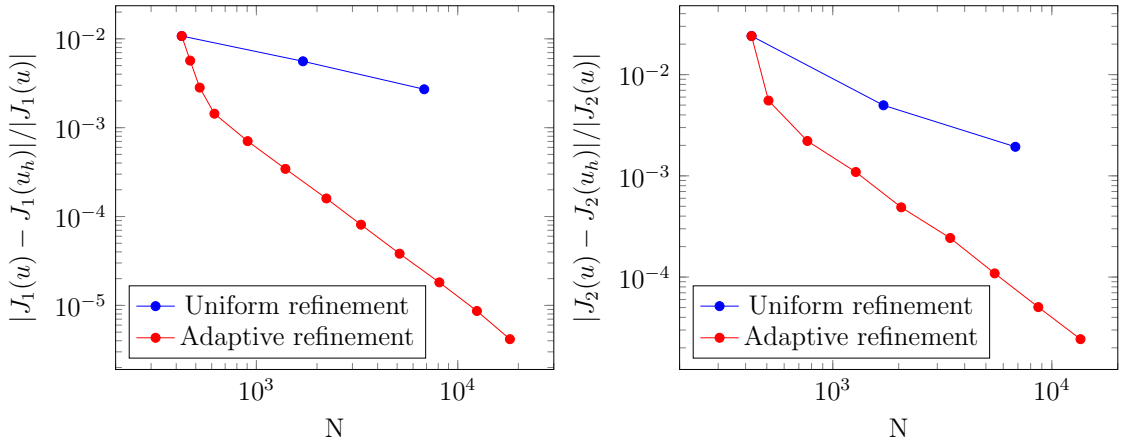

Finally in Figure \ref{fig:tongue graph2} we depict the efficiency {indices} for the global estimator $\eta_h$ and the sum of local estimators $\sum_K \eta_K$. 
{For both quantities $J_1$ and $J_2$, 
the two estimators provide an estimation of the \emph{discretization error} with an efficiency index around 1. In the case of $J_2$, we observe a slight overestimation for $\sum_K \eta_K$ and a slight underestimation for $\eta_h$.
}


\begin{figure}[H] 
\begin{center}
\begin{tikzpicture}[thick,scale=0.85, every node/.style={scale=1.0}] \begin{loglogaxis}[xlabel=N,
xmin=2e2,xmax=2e4,ymin=1e-1
,legend pos=south west, legend columns=1]
 \addplot[color=blue,mark=*] coordinates { 
(426.0,0.54751870568)
(469.0,0.600329745709)
(523.0,0.607981084252)
(619.0,0.628369525782)
(905.0,0.63019556358)
(1394.0,0.638589044224)
(2227.0,0.622205478868)
(3309.0,0.640799220961)
(5143.0,0.634907492951)
(8106.0,0.631988053072)
(12452.0,0.630752097027)
(18162.0,0.633745318834)
 };
  \addplot[color=red,mark=*]coordinates { 
(426.0,0.692697157585)
(469.0,0.808435456202)
(523.0,0.865543992835)
(619.0,0.889102947177)
(905.0,0.90578593186)
(1394.0,0.909334388886)
(2227.0,0.960593907027)
(3309.0,0.916151589076)
(5143.0,0.925914717127)
(8106.0,0.920538138678)
(12452.0,0.904397634638)
(18162.0,0.89903874333)
  };
 \legend{ $\eta_h/|J_1(u)-J_1(u_h)|$, $\sum_K\eta_K/|J_1(u)-J_1(u_h)|$ }
\end{loglogaxis} 
\end{tikzpicture} 
\begin{tikzpicture}[thick,scale=0.85, every node/.style={scale=1.0}] \begin{loglogaxis}[xlabel=N,
xmin=2e2,xmax=2e4,ymin=1e-1
,legend pos=south west, legend columns=1]
 \addplot[color=blue,mark=*] coordinates { 
(426.0,0.879211476907)
(509.0,0.70126372583)
(766.0,0.632538466558)
(1274.0,0.651580194578)
(2054.0,0.641165079058)
(3436.0,0.663994452488)
(5484.0,0.650999232161)
(8674.0,0.648138625808)
(13513.0,0.658281947168)
 };
  \addplot[color=red,mark=*]coordinates { 
(426.0,1.13684733872)
(509.0,1.41015753432)
(766.0,1.4260224086)
(1274.0,1.2071362381)
(2054.0,1.23989173483)
(3436.0,1.11961315105)
(5484.0,1.15236437714)
(8674.0,1.12208689261)
(13513.0,1.06089093943)
  };
 \legend{ $\eta_h/|J_2(u)-J_2(u_h)|$, $ \sum_K\eta_K/|J_2(u)-J_2(u_h)|$ }
\end{loglogaxis} 
\end{tikzpicture} 
\end{center}
\caption{Tongue model: efficiency indexes for $\eta_h$ (blue) and $\sum_K\eta_K$ \emph{vs.} the number $N$ of cells for the QoI $J_1$ (left) and $J_2$ (right).}\label{fig:tongue graph2}
\end{figure}
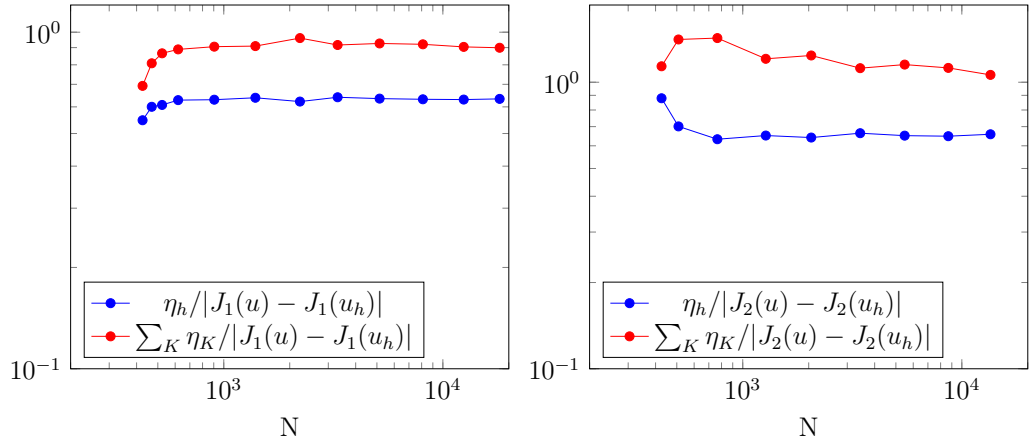

\subsection{Human artery with fiber activation}\label{sec:artery}

{As a second example we showcase the performance of the proposed algorithm for the analysis of the mechanical response of an artery with vulnerable coronary plaque to internal loading. 
Rupture of the cap induces the formation of a thrombus which may obstruct the coronary artery, cause an acute syndrome and the patient death.} The geometry (see Figure \ref{fig:artery medical img} (left)) comes from \cite{floch_vulnerable_2009} where the authors develop a methodology to reconstruct the thickness of the necrotic core area and the calcium area as well as the Young's moduli of the calcium, the necrotic core and the fibrosis. Their objective is the prediction of the vulnerable coronary plaque rupture.
{As represented in Figure \ref{fig:artery medical img} (left),
the diameter of the Fibrosis is equal to 5 mm.} 
{Following \cite{floch_vulnerable_2009}, we set different elastic parameters in each region: $E=0.011$ MPa, $\nu=0.4$ in the necrotic core and $E=0.6$ MPa, $\nu=0.4$ in the surrounding tissue.}
{No {volumetric} force field is applied: $\f=\0$.} 
We consider 
muscle fibers only in the media layer, where
smooth muscle cells are supposed to be perfectly oriented in the circumferential direction $\eb_A = \eb_\theta$, where $(\eb_r,\eb_\theta)$ is the basis for polar coordinates{{, see Figure \ref{fig:artery medical img} (center)}}. 
{Other parameters for fiber activation have been chosen as {$T = 0.01$ {MPa}} and $\beta=1$. 
}
As depicted in Figure \ref{fig:artery medical img} (right), the artery is fixed on the red portion {of} external boundary $\Gamma_D$. 
{Elsewhere, on the remaining part of the boundary, a homogeneous Neumann condition is applied: $\F=\0$. }
In the same figure, the green part represents the region of interest $\omega$. This choice is relevant in the study of vulnerable coronary plaque rupture.
{As in the previous example, we computed the relative displacement and the strain intensity, which maximal values are of 6.15 $\%$ and 0.3 $\%$, respectively. 
This ensures that small displacement and small strain assumptions are verified.
Figure \ref{fig:artery dual} 
depicts the magnitude of the solution in terms of displacements (left)
and the dual solutions associated to $J_1$ (center) and $J_2$ (right).
}

\begin{figure}[!h]
\begin{center}
\includegraphics[scale=0.80]{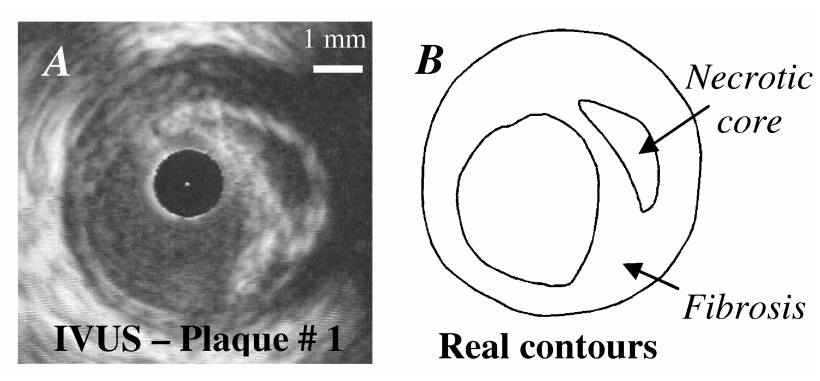} 
\includegraphics[scale=0.12]{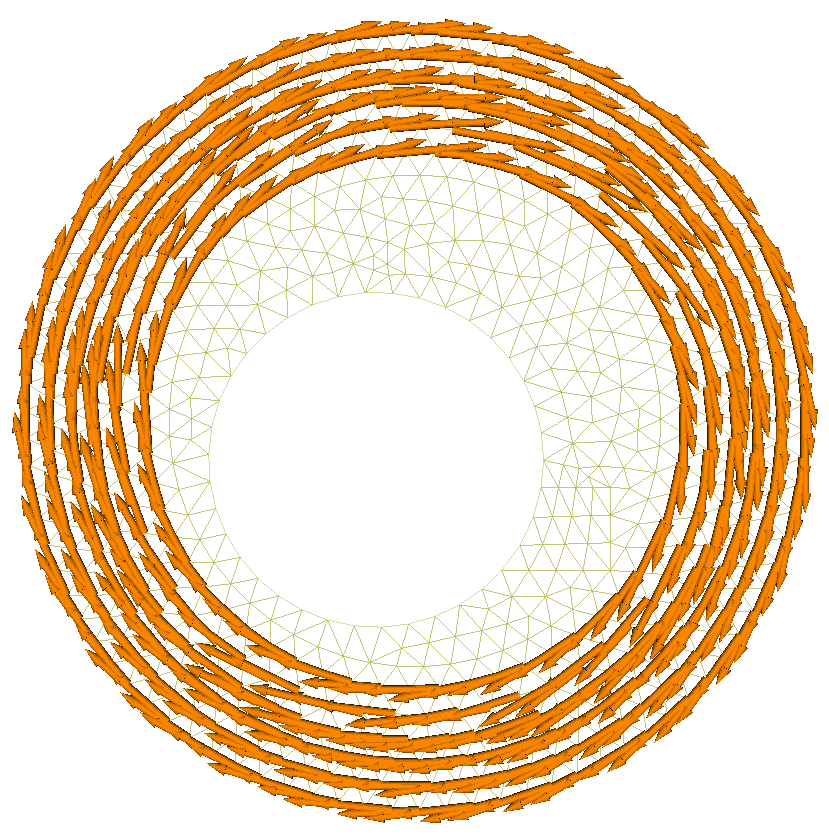} 
\includegraphics[scale=0.18]{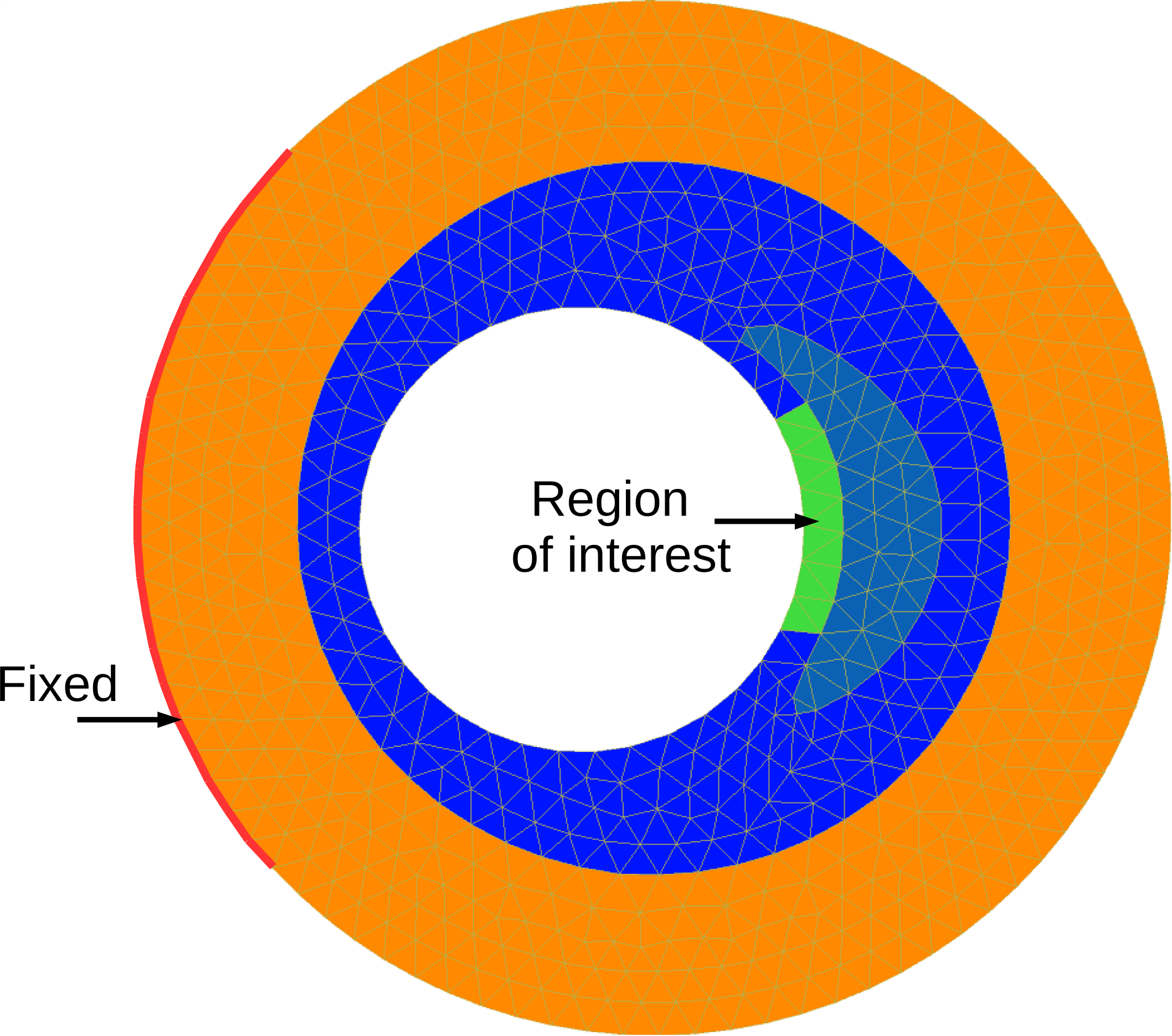}
\end{center}
\caption{Artery {model}: geometry (left), fiber orientation
 (center) and region of interest (right).}
\label{fig:artery medical img}\end{figure}




\begin{figure}[!h]
\begin{center}
\includegraphics[scale=0.08]{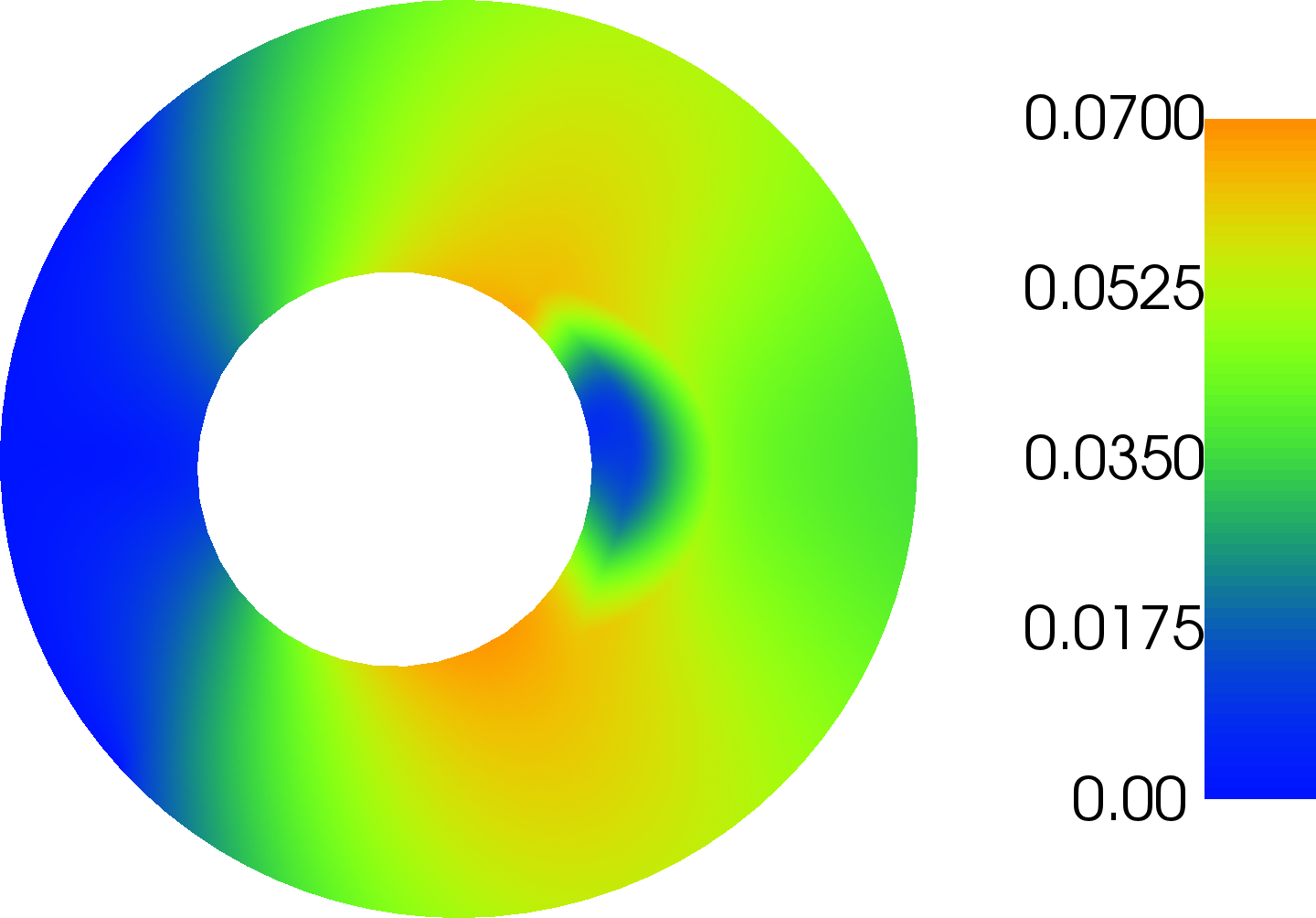}\hfill
\includegraphics[scale=0.08]{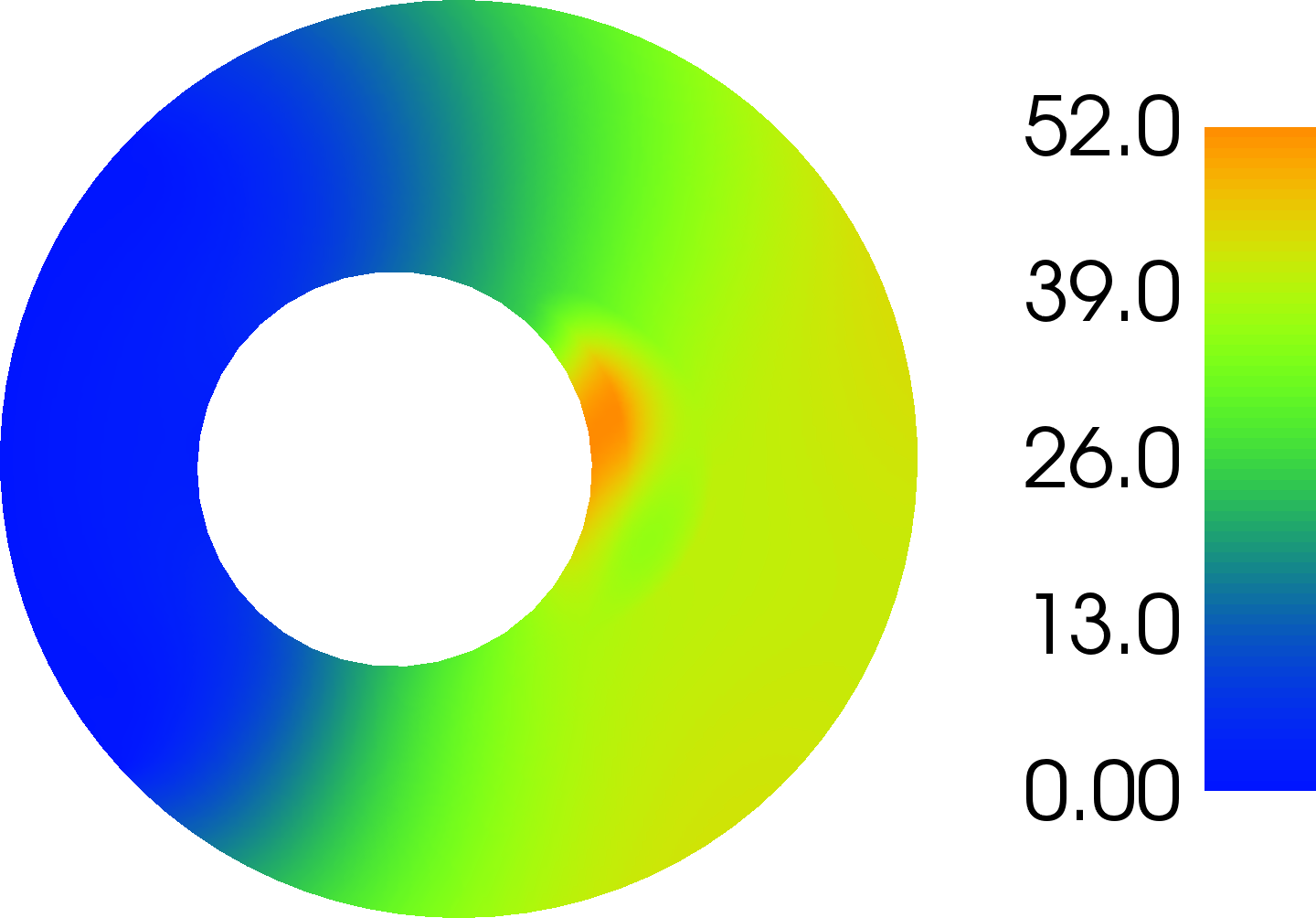}\hfill
\includegraphics[scale=0.08]{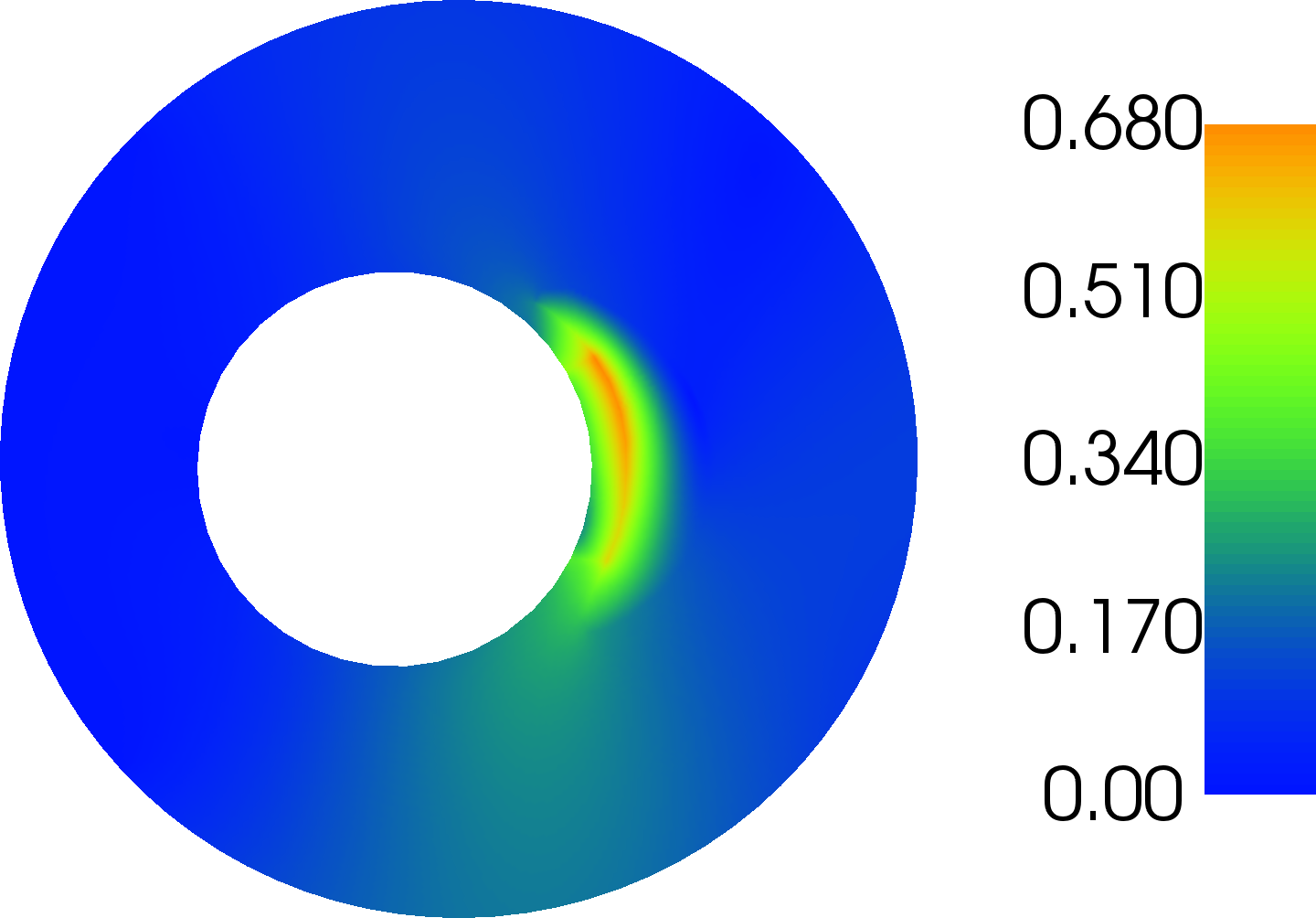}
\end{center}
\caption{Artery model: displacement (left), dual solution  for $J_1$ (center) and for $J_2$ (right).}
\label{fig:artery dual}\end{figure}
 
  

{
In Figure \ref{fig:artery meshes}, we present the final mesh after 2 and 6 iterations of Algorithm \ref{algo 1} for the quantity of interest $J_1$. 
As in the previous example, the refinement occurs in some specific regions{,} such as {those} near Dirichlet-Neumann transitions and concavities on the boundary. {Our results also show that the proposed method leads to}  the strong refinement near the interface between the necrotic core and the fibrosis, where stresses are localized because of the material heterogeneity. 
Conversely to the previous example, the refined meshes obtained for $J_2$ (not depicted) are very similar to those obtained for $J_1$.}



\begin{figure}[!h]
\begin{center}
\includegraphics[scale=0.119]{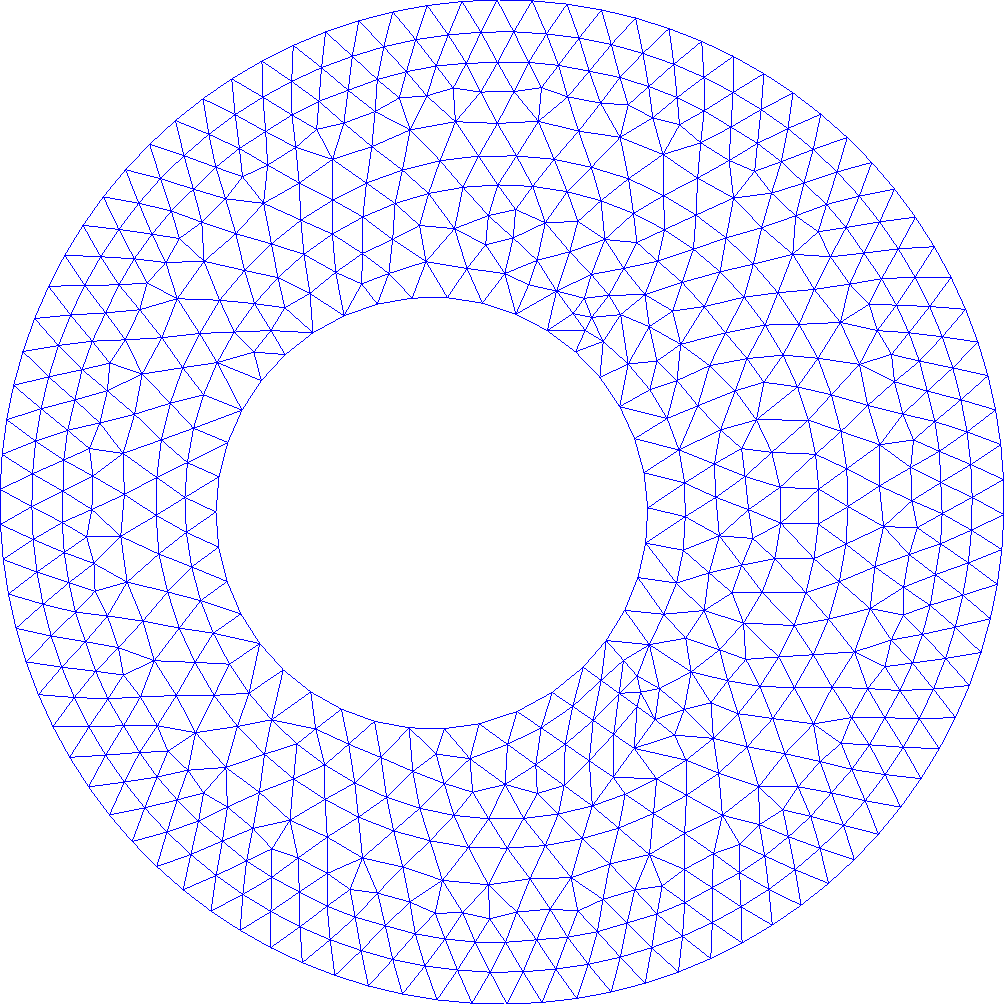}
\hspace{1em}
\includegraphics[scale=0.119]{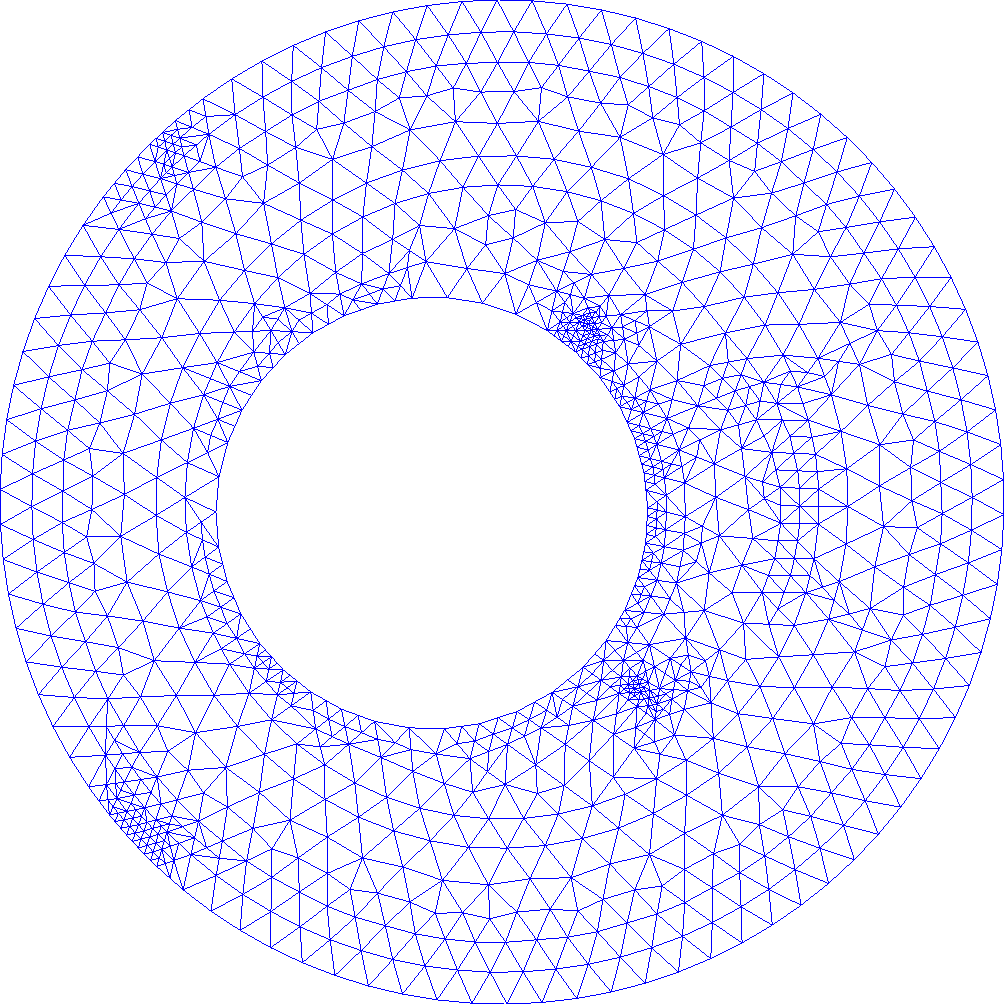}
\hspace{1em}
\includegraphics[scale=0.119]{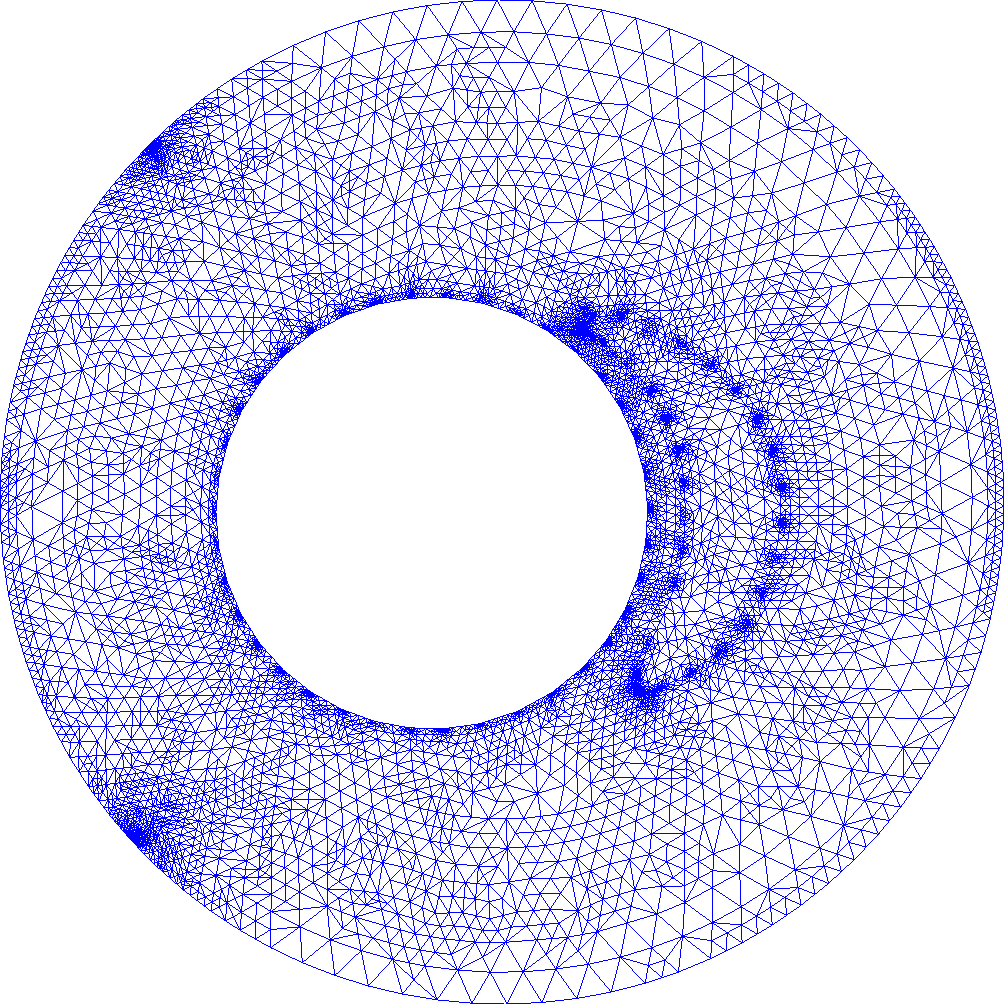} 


\end{center}
\caption{
Artery mesh: refinement driven by the QoI $J_1$.
Initial mesh (left) with 1242 cells and a relative error of 3.83e-1, adapted meshes after 2 iterations (center) with 2079 cells and a relative error of 5.25e-2 and after 6 iterations (right)
 with 15028 cells and a relative error of 3.37e-3.
 }
\label{fig:artery meshes}
\end{figure}

Figure \ref{fig:artery graph eff} (left) depicts the relative goal-oriented error 
$|J_1(\u) - J_1(\u_h)| / |J_1(\u)|$ 
versus the number $N$ of cells in the mesh, both for uniform refinement {(blue)} and adaptive refinement (red). 
{The stopping criterion $\varepsilon$ has been fixed at {5e-6}. 
{In Figure \ref{fig:artery graph eff} (right), we depict the efficiency {indices} for the global estimator $\eta_h$ and the sum of local estimators $\sum_K \eta_K$. 
The same observations as in the previous example can be stated, and
{the estimators provide acceptable value of the \emph{discretization error}. Moreover $\eta_h$ performs better though it still underestimates slightly the error.} Results we obtained for the quantity $J_2$ are very similar.

%

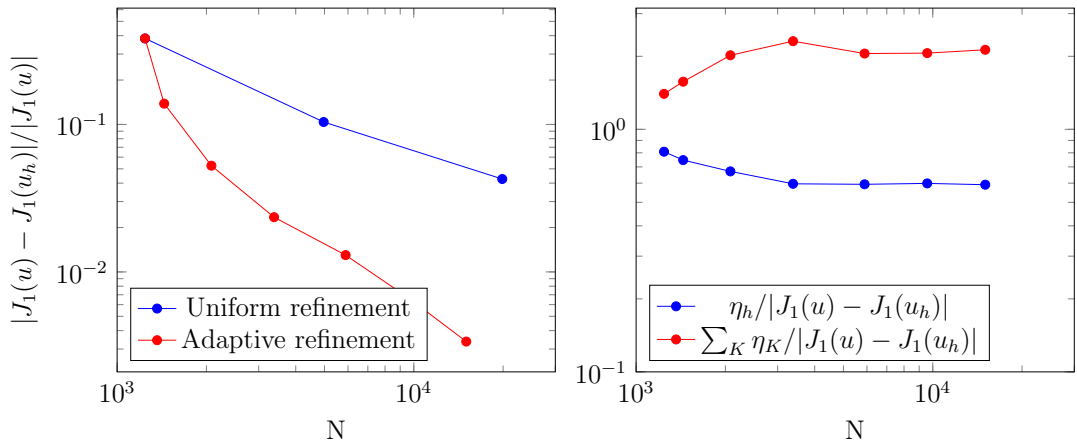
\begin{figure}[H] 
\begin{center}
\begin{tikzpicture}[thick,scale=0.85, every node/.style={scale=1.0}] \begin{loglogaxis}[xlabel=N,ylabel=$|J_1(u)-J_1(u_h)|/|J_1(u)|$,
xmin=1e3,xmax=3e4,xtickten={3,4}
,legend pos=south west, legend columns=1]
 \addplot[color=blue,mark=*] coordinates { 
(1242.0,0.383042850931)
(4968.0,0.103889179043)
(19872.0,0.0426608966597)
 };
  \addplot[color=red,mark=*]coordinates { 
(1242.0,0.383042850931)
(1441.0,0.138364267347)
(2079.0,0.0525334816088)
(3381.0,0.023503151137)
(5891.0,0.0130017151056)
(9578.0,0.00668885135909)
(15028.0,0.00337139785765)

 };
 \legend{ Uniform refinement, Adaptive refinement}
\end{loglogaxis} 
\end{tikzpicture} 
\begin{tikzpicture}[thick,scale=0.85, every node/.style={scale=1.0}] \begin{loglogaxis}[xlabel=N,
xmin=1e3,xmax=3e4,ymin=1e-1,xtickten={3,4}
,legend pos=south west, legend columns=1]
 \addplot[color=blue,mark=*] coordinates { 
(1242.0,0.808706493463)
(1441.0,0.746815968529)
(2079.0,0.67093292762)
(3381.0,0.596218349801)
(5891.0,0.593460833375)
(9578.0,0.598516436079)
(15028.0,0.591185139005)
 };
  \addplot[color=red,mark=*]coordinates { 
(1242.0,1.40028694858)
(1441.0,1.57197067518)
(2079.0,2.0195018399)
(3381.0,2.30936216451)
(5891.0,2.05327320072)
(9578.0,2.06154249942)
(15028.0,2.12780291125)
  };
 \legend{ $\eta_h/|J_1(u)-J_1(u_h)|$, $ \sum_K\eta_K/|J_1(u)-J_1(u_h)|$ }
\end{loglogaxis} 
\end{tikzpicture}
\end{center}
\caption{Artery {model}. Left: relative error for 
the QoI $J_1$ \emph{vs.} the number $N$ of cells in the case of uniform (blue) and adaptive (red) refinement.
Right: efficiency indexes of $\eta_h$ (blue) and $\sum_K\eta_K$ \emph{vs.} the number of cells $N$ for the QoI $J_1$. 
}
\label{fig:artery graph eff}
\end{figure}

\section{Discussion}\label{sec:discussion}

{In the first part, we discuss about the ability of the proposed methodology to assess and reduce the \emph{discretization error}. In a second part, we comment on some further issues to improve and guarantee the accuracy of the error estimator, and to optimize the mesh refinement algorithm. Finally we address the issue of tackling more complex problems that arise in current practice for clinical biomechanics, and point {out} the main limitations of the {current} study as well as some perspectives.}

\subsection{Towards quantification of the discretization error: first achievements}

The numerical results obtained in the last section show the ability of the proposed framework to provide relevant information about the \emph{discretization error}: though the global estimator $\eta_h$ provides only an approximation of the error in the quantity of interest $| J(\u) - J(\u_h) |$, this is often sufficient in practice. Moreover, the local estimators $\eta_K$ provide a means to evaluate ``relative'' errors and thereby drive mesh refinement (Algorithm \ref{algo 1}). Both the local and global errors can be significantly reduced without much computational effort. {For instance, in the first test-case \ref{sec:tongue}, and for $J_1$, the error is reduced by a factor of almost 4, after two successive refinements, and with only 20 $\%$ of extra cells.} In order to quantify more precisely the computational gains provided by the adaptive procedure, the computational time required to compute the error estimator and to regenerate or adapt the mesh should be thoroughly computed and analyzed, as was done for three-dimensional fracture problems treated by enriched finite element methods \cite{jin2017error}.  
%
%

Let us emphasize the well-known fact that sources of \emph{discretization errors} are \emph{local}, and concentrated mostly in regions where the solution is not smooth, e.g. subjected to strong variations, discontinuities or singularities. As a consequence, uniform refinement is highly suboptimal, while adaptive refinement performs much better by optimizing the number of elements, their size and location within the domain. 
Moreover, the proposed adaptive procedure is fully automatic, and no \emph{a priori} knowledge of the critical regions is needed. For goal-oriented error estimation,  the refined mesh obtained by the algorithm can in fact be counter-intuitive, because it is driven by the sensitivity of the quantity of interest with respect to the local error. This sensitivity is obtained by  solving the dual problem (see for instance Figure \ref{fig:tongue dual} in Section \ref{sec:tongue}) whose solution is, indeed, often not intuitive and difficult to interpret from a physical viewpoint.  

In comparison to widespread error techniques implemented in most of commercial finite element software, 
the DWR technique allows to estimate and {to} improve the error for an arbitrary quantity of interest $J$. Each practitioner can choose the relevant quantity {of interest $J$} 
and obtain an approximation of the error on this quantity of interest $| J(\u) - J(\u_h) |$, as well as a   map of the local error.  The authors emphasize that the results obtained in the current study also demonstrate that the optimal refinement strategy depends significantly on the choice of the quantity of interest $J$. In general, such a goal-oriented refinement strategy leads to meshes which may differ significantly from those obtained by minimizing the error in energy. 
Remark that such goal-oriented approaches were also developed for the Zienkiewicz-Zhu error estimators \cite{ZZ} in \cite{gonzalez2014mesh,gonzalez2015locally} and for explicit residual based estimates in \cite{ruter2013goal} and \cite{wick2016goal}. 

\subsection{Some further mathematical / computational issues}

It is desired that the global estimator $\eta_h$ provide reliable information on the error in the quantity of interest $| J(\u) - J(\u_h) |$, providing quality measures to the user. In theory, this error is a \emph{guaranteed upper bound}, with an explicit constant equal to $1$. Yet, the theory assumes that the dual solution $z$ is exactly known. This is never the case in practice as the dual problem is also solved using finite elements. Our numerical experiments show, however, that $| J(\u) - J(\u_h) |$ is estimated with reasonable accuracy and {that the} effectivity indices {are} close to 1, meaning that the approximate error on the quantity of interest is close to the (unknown) exact error on this quantity. 

The numerical experiments provided in this paper confirm those of the literature on DWR technique, \emph{e.g.}, \cite{becker-rannacher-2001,giles-suli-2002,rognes-logg-2013} {showing that the DWR is, in most situations, a reliable approach to compute goal-oriented error estimates}.
{However, in certain situations, the DWR estimator is not as reliable as desired, since the effect of approximating the dual solution is difficult to control. This issue has been already pointed in the literature: see {\emph{e.g.} \cite{carstensen-2005,nochetto-veeser-verani-2009,ainsworth-rankin-2012} and earlier considerations in, \emph{e.g.}, \cite{giles-suli-2002,bangerth-rannacher-2003}. 
Especially, in \cite{nochetto-veeser-verani-2009} a simple situation where $\eta_h$ provides a poor estimation on a coarse mesh is detailed. 
There is up to now no simple, cheap and general technique to address this issue, but first solutions have been suggested in \cite{carstensen-2005,nochetto-veeser-verani-2009,ainsworth-rankin-2012}. They consist in modifying the DWR estimator so as to take into account the approximation of $z$. This is a stimulating perspective for further research. Moreover, the issue of computing a cheaper approximation of $z$, without compromising the reliability and efficiency of the estimator still needs to be addressed in depth.}

Concerning mesh refinement, though the local estimator $\eta_K$ combined with Algorithm \ref{algo 1} provides acceptable results, no effort has been spent on finding the value of  parameter $\alpha$ in the D\"orfler marking that yields improved refined meshes. On this topic, our global strategy for error estimation and mesh refinement is only a first attempt, and can be improved. For instance, in \cite{BET11}, an adaptive method based on specific weighting of the residuals of the primal and dual problems has been designed, and leads to quasi-optimal adapted meshes. Such a method could be tested and compared to the current one.

\subsection{Applicability for patient-specific biomechanics?}

Though the preliminary results presented in this paper demonstrate the relevance and practicability of  a posteriori error estimators for providing quality control in quantities of interest to the biomechanics practitioner, and to drive mesh adaptation, much effort is still needed for the approaches developed here to address practical, personalized, clinically-relevant Finite Element (FE) simulations for biomechanical applications.

First, the compressible linear framework considered here is inadequate in practice and must be replaced by a fully non-linear, incompressible, time and history dependent model \cite{payan-ohayon-2017}.
Non-linearities {also} occur due to boundary conditions, when, for instance, contact is present \cite{courtecuisse2014real}. Moreover, most of the quantities of interest in biomechanics are non-linear (norm of the displacements, local shear stress, {maximum admissible stress and strain, }etc). {It is important to point out here} that the DWR method for goal-oriented error estimation is already capable of tackling non-linearities: see, \emph{e.g.}, \cite{becker-rannacher-2001} for the general framework, and, \emph{e.g.}, \cite{larsson-hansbo-runesson-2002,whiteley-tavener-2014} for first applications in non-linear elasticity and \cite{ruter2013goal} for fracture mechanics. Nevertheless, this non-linear framework needs to be adapted and tested in the specific case of hyperelastic soft-tissue. 

{
The major limitation of our work is that it assumes that the mathematical model used to describe the biomechanics problem is able to reproduce the physical reality. Unfortunately, in general, selecting the proper mathematical model for a given biomechanics problem is probably the most challenging part of the simulation process. The large, and increasing, number of papers dealing with the choice of constitutive model, for example, testifies for this difficulty. 
For a wide range of problems, indeed, modeling errors are the most significant. Estimating rigorously and systematically the impact of these errors is extremely challenging, in particular when dealing with patient-specific simulations. Dealing with this issue is the focus of ongoing research in our 
teams but is far beyond the scope of this paper. 

We would nonetheless like to make the following remarks. The first problem which must be addressed is the choice of a model (hyperelastic, viscous, porous, single/multi-scale...). The chosen model has parameters which must be estimated through inverse analysis. Once estimates, or probability distributions for these parameters are available, their importance on quantities of interest must be evaluated, through sensitivity analysis and uncertainty quantification. 
The major difficulty is, therefore, to select the proper model, and its parameters for a given patient. As in vivo experiments are in general not possible, data must be extracted as the patient is being treated, e.g. during an operation. This can be done using Bayesian methods, which provide a reconciliation between expert knowledge on patient cohorts (prior) and actual properties of a given patient \cite{rappel2016bayesian,rappel2017bayesian}. Real-time machine-learning-like methods such as Kalman filters 
demonstrated as well promising results \cite{moireau2009,haouchine2013image}. To evaluate the effects of uncertainties on such material parameters, accelerated Monte-Carlo methods are possible avenues of investigation \cite{hauseux2017accelerating}. 
An exciting question is the comparative usefulness and combination of physical models (potentially learnt during medical treatment) and machine-learning algorithms, mostly based on data acquired during the intervention.
Last but not least, note that the DWR method is based on optimal control principles, that makes it suitable for extensions to parameter calibration (viewed as an optimal control problem). In such a setting, it allows to combine sensitivity analysis with goal-oriented {\it a posteriori} error estimation, see \cite{becker-vexler-2005}. In the same spirit, the interplay between {\it a posteriori} error estimation and uncertainty quantification has been object of recent research interests \cite{eigel-2016,guignard-2016}.

We also note that if users can obtain some estimate, even rough, of modeling errors, they will {also} be able to compare discretization and model errors. This enables the coarsening of the mesh if the discretization error is unnecessarily small in comparison to the modeling error as is done, e.g. in \cite{akbari2015scale} and \cite{talebi2013molecular,budarapu2014efficient,talebi2014computational,silani2014semi,talebi2015concurrent} for adaptive scale selection. 
Conversely, for specific applications where modeling errors are small or moderate, the mesh can be refined efficiently to increase the precision.}

With our methodology, practitioners spending a large amount of time and effort in patient-specific mesh generation can obtain useful information on the impact of the quality of the mesh on quantities of interest to them. This information goes well beyond purely geometrical criteria for the regularity of the elements which are typically provided in commercial software. 

This information can be used directly to optimize the choice of the discretization/mesh in view of minimizing the error on a specific quantity of interest. Fast/real-time numerical methods which provide  real-time predictions have been intensively researched since the beginning of the 1990's. Those approaches are critical to build surgical planning and guidance tools, for example. Reliable error estimation is critical in these situations to guarantee the accuracy, but has been extremely scarcely addressed in the literature.
As a first step in this direction, the recent work of  \cite{bui2017real} provides a real-time mesh refinement algorithm for needle insertion. Mesh refinement is driven by a ZZ error estimate, for the global norm. It would be interesting to extend such a method for goal-oriented error estimation, e.g. on the motion of a target, or reaction/friction force along the needle shaft. 

We should also mention {alternative approaches to (implicit, standard) finite elements for fast nonlinear finite element analysis: for instance the solution of total lagrangian formulation of the equilibrium equations on graphics processing unit for neurosurgical simulation \cite{joldes_suite_2009,joldes_non-locking_2009,joldes_real-time_2010},  or model order reduction techniques for the real-time, interactive simulation of tissue tearing during laparoscopic surgery \cite{niroomandi_real-time_2012,niroomandi_real-time_2013,quesada_haptic_nodate}. A  perspective consists in extending the current framework to such numerical methods where error control is particularly challenging. For explicit approaches, the interplay between the choice of the time-step and that of the mesh size is a difficult topic, especially for domains with significant stiffness differences where adaptive and multi-time-step schemes should be investigated.}  

\section*{Acknowledgements}
The authors thank warmly Mathias Brieu 
for the organization of the EuroMech 595 workshop. They also thank Roland Becker, Jean-Louis Martiel, Jacques Ohayon and Yohan Payan for their support and helpful comments that allowed to improve the paper, as well as the following people for inspiring discussions: Jack Hale, Florence Hubert and Pascal Perrier.
For funding, F. Chouly thanks R\'egion Bourgogne Franche-Comt\'e (``Convention R\'egion 2015C-4991. Mod\`eles math\'ematiques et m\'ethodes num\'eriques pour l'\'elasticit\'e non-lin\'eaire''), the Centre National de la Recherche Scientifique (``Convention 232789 DEFI InFIniTI 2017 - Projet MEFASIM'')
and the Agence Maths Entreprises (AMIES) (``Projet Exploratoire PEPS2 MethASim''). 




\section*{\refname}

\bibliographystyle{elsarticle-harv}
\bibliography{biblio}





\end{document}